\documentclass[11pt, a4paper, final]{article}

\pdfoutput=1

\usepackage[english]{babel}
\usepackage[utf8]{inputenc}
\usepackage{pdflscape}
\usepackage{enumerate}
\usepackage{amsbsy}
\usepackage{amsmath} 
\usepackage{graphics}
\usepackage{mathrsfs}
\usepackage{wrapfig}
\usepackage{placeins}
\usepackage{mathtools}

\usepackage{amsfonts}
\usepackage{pstricks}
\usepackage{setspace}

\usepackage{showlabels}
\makeatletter
\showlabels{blx@bibitem}
\makeatother

\usepackage[babel]{csquotes}
\usepackage[subentry,backend=biber,sorting=none,style=numeric-comp,natbib=true]{biblatex}
\newbibmacro*{bbx:parunit}{%
  \ifbibliography
    {\setunit{\bibpagerefpunct}\newblock
     \usebibmacro{pageref}%
     \clearlist{pageref}%
     \setunit{\adddot\par\nobreak}}
    {}}

\renewbibmacro*{doi+eprint+url}{
  \iftoggle{bbx:doi}
    {\iffieldundef{eprint}{\printfield{doi}}{}}
    {}%
  \iftoggle{bbx:eprint}
    {\usebibmacro{eprint}}
    {}%
  \iftoggle{bbx:url}
    {\iffieldundef{eprint}{\iffieldundef{doi}{\usebibmacro{url+urldate}}{}}{}}
    {}}

\renewbibmacro*{eprint}{%
  \iffieldundef{eprinttype}
    {\printfield{eprint}}
    {\printfield[eprint:\strfield{eprinttype}]{eprint}}}

\renewbibmacro*{url+urldate}{%
  \printfield{url}%
  \iffieldundef{urlyear}
    {}
    {\setunit*{\addspace}%
     \printtext[urldate]{\printurldate}}}

\makeatletter
\DeclareFieldFormat{eprint:arxiv}{%
  arXiv\addcolon~\ifhyperref
    {\href{http://arxiv.org/\abx@arxivpath/#1}{\nolinkurl{#1}}}{\nolinkurl{#1}}%
    \iffieldundef{eprintclass}{}{\nobreakspace\texttt{\mkbibbrackets{\thefield{eprintclass}}}}}
\makeatother

\DeclareFieldFormat{doi}{\mkbibacro{doi}\addcolon~\ifhyperref
  {\href{http://dx.doi.org/#1}{\nolinkurl{#1}}}
  {\nolinkurl{#1}}}

\DeclareFieldFormat{url}{\mkbibacro{url}\addcolon~\url{#1}}

\patchcmd{\bibsetup}{\interlinepenalty=5000}{\interlinepenalty=10000}{}{}

\DeclareFieldFormat[article]{title}{\textit{#1}}
\DeclareFieldFormat[article]{journaltitle}{#1}
\DeclareFieldFormat[article]{volume}{\textbf{#1}}
\DeclareFieldFormat[article]{number}{\textbf{#1}}
\DeclareFieldFormat[article]{pages}{#1}
\renewbibmacro{in:}{}

\appto{\bibsetup}{\raggedright}

\defbibentryset{inflation}{Kazanas:1980,Guth:1980,Sato:1980,Mukhanov:1981,Linde:1981,Albrecht:1982,Hawking:1981,Chibisov:1982,Hawking:1982,Guth:1982,Starobinsky:1982}
\defbibentryset{SM}{Sasaki:1986,Mukhanov:1988}

\makeatletter
\newcommand{\dontusepackage}[2][]{
  \@namedef{ver@#2.sty}{9999/12/31}%
  \@namedef{opt@#2.sty}{#1}}
\makeatother

\dontusepackage[numbers,sort&compress]{natbib}		

\usepackage{jcappub}					

\expandafter\let\csname ver@natbib.sty\endcsname\relax
\expandafter\let\csname opt@natbib.sty\endcsname\relax

\usepackage[noabbrev]{cleveref}			

\newcommand{\bea}{\begin{eqnarray}} \newcommand{\eea}{\end{eqnarray}}
\newcommand{\el}{\nonumber \\}
\newcommand{\re}[1]{(\ref{#1})}

\newcommand{\pat}{\partial}

\renewcommand{\sec}[1]{section \ref{#1}}
\newcommand{\fig}[1]{figure \ref{#1}}

\newcommand{\para}{\paragraph}

\renewcommand{\a}{\alpha}
\renewcommand{\b}{\beta}
\renewcommand{\c}{\gamma}
\renewcommand{\d}{\delta}

\renewcommand{\l}{\lambda}

\newcommand{\LCDM}{$\Lambda$CDM\ }

\newcommand{\ha}{\frac{1}{2}}

\newcommand{\rmd}{\mathrm{d}}

\newcommand{\ie}{i.e.\ }

\newcommand{\half}{\ha}

\newcommand{\kappat}{\bar{\kappa}}
\newcommand{\xh}{x_{\mathrm{max}}}
\newcommand{\deh}{\d_{x}}

\newcommand{\sR}{{\mathcal{R}}}

\newcommand{\Mpl}{M_{\mathrm{Pl}}}
\newcommand{\rme}{{\mathrm{end}}}

\newcommand{\reh}{{\mathrm{reh}}}

\usepackage{soul}

\title{Higgs inflation with loop corrections in the Palatini formulation}

\author[a,b]{Syksy R\"{a}s\"{a}nen}

\author[a]{and Pyry Wahlman}

\affiliation[a]{University of Helsinki, Department of Physics
and Helsinki Institute of Physics \\
P.O. Box 64, FIN-00014 University of Helsinki, Finland}

\affiliation[b]{Kobe University, Department of Physics, Kobe 657-8501, Japan}

\emailAdd{syksy.rasanen@iki.fi}

\emailAdd{pyry.wahlman@helsinki.fi}

\abstract{
We compare Higgs inflation in the metric and Palatini formulations of general relativity, with loop corrections treated in a simple approximation. We consider Higgs inflation on the plateau, at a critical point, at a hilltop and in a false vacuum. In the last case there are only minor differences.
Otherwise we find that in the Palatini formulation the tensor-to-scalar ratio is consistently suppressed, spanning the range $1\times10^{-13}<r<7\times10^{-5}$, compared to the metric case result $2\times10^{-5}<r<0.2$. Even when the values of $n_s$ and $r$ overlap, the running and running of the running are different in the two formulations. 
Therefore, if Higgs is the inflaton, inflationary observables can be used to distinguish between different gravitational degrees of freedom, in this case to determine whether the connection is an independent variable. Non-detection of $r$ in foreseeable future observations would not rule out Higgs inflation, only its metric variant. We conclude that in order to fix the theory of Higgs inflation, not only the particle physics UV completion but also the gravitational degrees of freedom have to be explicated.
}

\begin{document}

\begin{flushleft}
	\hfill		 HIP-2017-23/TH \\
	\hfill		 KOBE-COSMO-17-12
\end{flushleft}

\maketitle
  
\setcounter{tocdepth}{2}

\setcounter{secnumdepth}{3}

\section{Introduction} \label{sec:intro}

\para{Extending the Standard Model to high scales and Higgs inflation.}

The range of validity of the Standard Model of particle physics has proven to be wider than expected, as no physics beyond it has been discovered in collider experiments. Indeed, the only new piece of information (stronger exclusion limits on extensions of the Standard Model aside) that the LHC has uncovered is the Higgs mass, $m_H=125.09\pm0.21\pm0.11$ GeV \cite{Aad:2015}. It is noteworthy that this value is such that the Standard Model may be a viable theory all the way to the Higgs field taking Planck scale values $\Mpl=2.4\times10^{18}$ GeV, or even beyond (the energy density remains sub-Planckian at Planck scale field values). The main limit comes from top mass loop contributions possibly driving the quartic Higgs self-coupling negative at large field values, making the electroweak vacuum unstable. The stability limit is sensitive to the precise values of the Higgs and top masses and the strong coupling constant, and present values are consistent with stability, instability and metastability at the Planck scale, within the experimental and theoretical errors \cite{Espinosa:2015a, Espinosa:2015b, Iacobellis:2016, Butenschoen:2016, Espinosa:2016}.
The fact that the Standard Model sits close to the instability limit (\ie that the quartic Higgs coupling runs close to zero) at the Planck scale is an unexpected feature. Such behaviour was used to predict the value of the Higgs mass in the context of asymptotic safety \cite{Shaposhnikov:2009}. Validity of the Standard Model to high scales also means that Higgs could possibly be the inflaton \cite{Bezrukov:2007, Bezrukov:2013}.\footnote{As we discuss below, in this case it is not clear what is the correct way to treat the loop corrections, so stability at large field values is even more unsettled.}

In cosmology, inflation has become the standard scenario for the primordial universe. It alleviates the homogeneity and isotropy problem \cite{Trodden:1998, Trodden:1999, Ellis:1999, Ellis:2000, East:2015}, explains spatial flatness and has predicted in detail that initial perturbations are predominantly adiabatic, close to scale-invariant, highly Gaussian and predominantly scalar \cite{Starobinsky:1980, inflation, SM}, in excellent agreement with observations \cite{Planck:inflation}.\footnote{As there are inflationary models where some of these properties do not hold, calling them predictions may be too strong, but they are certainly generic features of inflation.} If the non-minimal coupling of the Higgs field to gravity is neglected, then at large field values the potential can be approximated as $\frac{1}{4}\l(h)h^4$, where the change in the potential due to loop corrections is taken into account as running of the quartic coupling with the field. The potential is not sufficiently flat to allow a long enough period of inflation with a small enough amplitude of perturbations \cite{Isidori:2007, Hamada:2013, Fairbairn:2014}.

However, if the Higgs field is non-minimally coupled to gravity, the inflationary behaviour changes. Even if the non-minimal coupling $\xi$ is not input at the classical level, it will be generated by quantum corrections \cite{Callan:1970}. Even if $\xi$ is set to zero on some given scale, it will be non-zero on other scales due to renormalisation group running.
Inflation with non-minimal coupling has been considered since the 1980s \cite{Futamase:1989, Salopek:1989, Fakir:1990, Makino:1991, Fakir:1992, Kaiser:1994, Komatsu:1999}, but Higgs inflation is a remarkably simple model, which uses the only known scalar field that may be elementary instead of composite, and does not introduce any new parameters that are not required by the theory.
If only the tree level Standard Model Higgs potential is considered, the non-minimal coupling makes the potential flat at high field values, and the inflationary predictions for the spectral index and the tensor-to-scalar ratio depend only on the number of e-folds until the end of inflation \cite{Bezrukov:2007}. As the Standard Model field content and couplings are precisely known\footnote{Though dark matter and baryogenesis physics, at least, has to be added. The $\nu$MSM is an economical possibility \cite{Asaka:2005a, Asaka:2005b, Bezrukov:2008a, Bezrukov:2011a}, where Higgs inflation is unchanged.}, in Higgs inflation the processes of preheating and reheating can be calculated in detail, removing the uncertainty in the number of e-folds that accompanies less complete embeddings of inflation into particle physics \cite{Bezrukov:2008a, GarciaBellido:2008, Figueroa:2009, Figueroa:2015, Repond:2016}\footnote{This may also make it possible to distinguish Higgs inflation from $R^2$ inflation \cite{Starobinsky:1980, Bezrukov:2011b}, where the slow-roll stage is very similar.} (see also \cite{Ema:2016}). Higgs decay also produces a distinctive signature of gravitational waves (also present, though different, when Higgs is not the inflaton) \cite{Figueroa:2014}. The tree level predictions agree well with observations \cite{Planck:inflation}.

Quantum corrections complicate the picture. The loop modifications to the potential, and thus the inflationary predictions, depend on the Higgs and top mass measured at low energies, which offers a novel consistency test between collider experiments and cosmological observations \cite{Espinosa:2007, Barvinsky:2008, Burgess:2009, Popa:2010, DeSimone:2008, Bezrukov:2008b, Bezrukov:2009, Barvinsky:2009, Bezrukov:2010, Bezrukov:2012, Allison:2013, Salvio:2013, Shaposhnikov:2013}.
However, as the Standard Model coupled to gravity is non-renormalisable, it is not clear how the loop corrections should be calculated, and the relation between the low energy and high energy regime is still debated \cite{George:2013, Calmet:2013, Bezrukov:2014a, Bezrukov:2014b, Rubio:2015, George:2015, Saltas:2015, Fumagalli:2016, Enckell:2016, Bezrukov:2017, Fumagalli:2016sof}.
At low energies, where gravity can be neglected, the Standard Model is renormalisable, and at high energies there is an approximate shift symmetry that keeps loop corrections small, but the matching between the two regimes depends on the ultraviolet completion of the theory \cite{Bezrukov:2010, George:2013, Bezrukov:2014b, George:2015, Fumagalli:2016, Enckell:2016, Bezrukov:2017, Fumagalli:2016sof}. Put another way, because the theory is non-renormalisable, the low energy theory does not uniquely fix the predictions and the results depend on the renormalisation prescription.

It has been claimed that the large non-minimal coupling $\xi$ lowers the scale of perturbative violation of unitarity from the Planck scale down to the energy $\sim\Mpl/\xi$ \cite{Barbon:2009, Burgess:2009, Hertzberg:2010}.\footnote{For the Palatini formulation discussed below, naive power counting gives the scale $\Mpl/\sqrt{\xi}$ instead \cite{Bauer:2010}.} However, it has been countered that the cutoff scale should depend on the Higgs field \cite{Bezrukov:2010, Bezrukov:2011a}, and when this is taken into account, the theory is during inflation perturbatively valid at the scale relevant for inflation. In any case, no explicit violation of unitarity in the inflationary regime has been demonstrated. It has been argued using resummation \cite{Calmet:2013} that, in contrast to naive power-law counting arguments, the theory remains unitary all the way to the Planck scale.\footnote{It has been claimed that this follows from writing the action in terms of gauge-invariant variables, but only tree-level 2-to-2 Higgs scattering has been considered so far \cite{Weenink:2010, Prokopec:2012, Prokopec:2014}.}
It has also been claimed that the treatment ceases to be valid for Higgs field values (as opposed to energies) larger than $\Mpl/\xi$, as higher order corrections spoil the flatness of the potential required for inflation \cite{Barbon:2009} and alter the renormalisation group running of the quartic Higgs coupling \cite{Burgess:2014}.
If a theory ceases to be valid on some scale, then using an expansion with terms suppressed by that scale is not expected to work close to that scale, and definite statements would require knowing the ultraviolet completion \cite{Espinosa:2015b}.
As the completion of Higgs inflation, which includes the Standard Model and general relativity, is unknown, we do not know whether such terms are present or not. There is not necessarily any problem in using field values larger than the Planck scale, as long as the energy density remains sub-Planckian (see section 2.4 of \cite{Linde:2005}), as is the case in Higgs inflation.

The loop corrections not only affect the mapping between the inflationary parameters and low energy observables, they can also open qualitatively new inflationary regimes by changing the shape of the potential. In addition to inflation on the flat plateau originally proposed, it is possible that the Higgs field drives inflation at an inflection point (or an almost inflection point, called a critical point) \cite{Allison:2013, Bezrukov:2014a, Hamada:2014, Bezrukov:2014b, Rubio:2015, Fumagalli:2016, Bezrukov:2017}, on the hilltop \cite{Fumagalli:2016} or in a degenerate vacuum in a hillclimbing scenario \cite{Jinno:2017a, Jinno:2017b}. It is also possible to have false vacuum Higgs inflation, although in that case new physics is needed to end up in the electroweak vacuum, one suggestion involving non-minimal coupling to gravity \cite{Masina:2011a, Masina:2011b, Masina:2012, Masina:2014, Notari:2014, Fairbairn:2014, Iacobellis:2016}. False vacuum and critical point Higgs inflation have the interesting property that they can produce a sizeable tensor-to-scalar ratio $r\sim0.1$, unlike plateau inflation, which gives (at tree level) $r=5\times10^{-3}$.

\para{Palatini gravity.}

If Higgs is the inflaton, then in addition to (at least in principle) connecting inflationary observables to electroweak scale particle physics, it also opens a window on the theory of gravity. In the usual formulation of general relativity, called the metric formulation, the independent variables are the metric and its first derivative. Even though the action involves the second derivative of the metric, the equations of motion obtained by varying it are nevertheless only second order due to the unusual structure of the Einstein--Hilbert action \cite{Padmanabhan:2004, Paranjape:2006}. (In higher dimensions, this property is shared by the other Lovelock Lagrangians \cite{Lovelock:1971, Lovelock:1972}.) However, deriving the equations of motion requires adding the York--Gibbons--Hawking boundary term to the action to cancel a total derivative term that depends on the first derivatives of the metric, the variation of which cannot be taken vanish on the boundary \cite{York:1972, Gibbons:1977}.

In the Palatini formulation of general relativity, the independent variables are instead taken to be the metric and the connection.\footnote{The formulation is usually credited to the paper \cite{Palatini:1919} by Palatini, but it appears to have been first presented in the paper \cite{Einstein:1925} by Einstein \cite{Ferraris:1982}.} The metric thus appears as an auxiliary variable with no kinetic term. The variation of the action with respect to the metric gives the Einstein equation, and variation with respect to the connection determines the connection in terms of the metric.
For Lovelock Lagrangians, the Palatini formulation leads to the same equations of motion as general relativity and to the same, namely Levi--Civita, connection, and is thus physically equivalent, at least at the classical level \cite{Exirifard:2007}. Using the Palatini formulation is not an extra assumption about the theory, just a different parametrisation of the gravitational degrees of freedom. It could be argued that the Palatini formulation is simpler than the metric formulation, because there is no need to add a boundary term to the action, as it involves only first derivatives of the variables. The first order formulation is also more natural from the point of view of canonical quantisation. For actions more general than the Lovelock action, the metric and the Palatini formulation are physically inequivalent. In particular, this is true when the Higgs is non-minimally coupled to gravity. 

Thus, in addition to the particle physics ambiguity resulting from sensitivity to the particle physics UV completion, there is also another ambiguity, related to choosing the gravitational degrees of freedom (which presumably are determined by the UV completion of the gravitational part of the theory). Turning this around, if Higgs is the inflaton, inflationary observables can be used as probes of the gravitational degrees of freedom.

The difference in the predictions of the metric and Palatini formulations was worked out in \cite{Bauer:2008} for Higgs plateau inflation at tree level. We extend the analysis to include loop corrections, treated in a simplified way, and consider also inflation outside of the plateau. In \sec{sec:ana} we discuss plateau, critical point, hilltop and false vacuum inflation analytically. In \sec{sec:num} we do a numerical calculation to fully cover the parameter space, and compare to the analytical results and previous work. We summarise our results in \sec{sec:conc}.

\section{Analytical study} \label{sec:ana}

\subsection{The potential and the two frames} \label{sec:pot}

\para{Action.}

At the classical level, we have the usual Standard Model action plus the Einstein--Hilbert action, along with a term that couples the Higgs doublet directly to the Ricci scalar $R$. We consider only the radial Higgs mode $h$, which corresponds to the physical Higgs field (for the effects of the three other components, the would-be Goldstone modes, see \cite{Hertzberg:2010, Mooij:2011, Greenwood:2012, George:2013, George:2015}), so the relevant part of the action is, in the non-minimally coupled Jordan frame,
\bea \label{actionJ}
  S = \int\rmd^4 x \sqrt{-g} \left( \frac{M^2 + \xi h^2}{2} g^{\a\b} R_{\a\b}(\Gamma,\pat\Gamma) - \frac{1}{2} g^{\a\b} \nabla_\a h \nabla_\b h - V(h) \right) \ ,
\eea
where $h$ is the Higgs field with the potential $V(h)=\frac{\l}{4}(h^2-v^2)^2$.
In our sign convention $\xi=-\frac{1}{6}$ corresponds to the conformally coupled case; we only consider positive values of $\xi$, as negative values are not suitable for Higgs inflation.
The Planck mass today is $\Mpl=\sqrt{M^2+\xi v^2}=2.4\times10^{18}$ GeV. The limit from the LHC is $|\xi|<2.6\times 10^{15}$ \cite{Atkins:2012}, so as $v=246$ GeV, we have $\Mpl\approx M$; we use units such that $M=1$. In the action \re{actionJ}, $\Gamma$ is the connection. If we assume that $\Gamma$ is the Levi--Civita connection, we have the metric formulation, whereas keeping $\Gamma$ as an independent variable corresponds to the Palatini formulation. Note that the equation of motion for $h$ involves the covariant derivative with respect to the Levi--Civita connection even in the Palatini formulation. (This is determined by the requirement that when varying the action a total derivative term that involves the variation of $h$ becomes a boundary term that vanishes.)

The action can be brought into the non-minimally coupled form by writing $g_{\a\b}=\Omega^{-2}\tilde g_{\a\b}\equiv(1+\xi h^2)^{-1}\tilde g_{\a\b}$. This field redefinition is the same in both the metric and the Palatini formulation, and it multiplies the kinetic term of $h$ with the factor $(1+\xi h^2)^{-1}$. In the metric formulation, $R_{\a\b}$ depends on $g_{\a\b}$, resulting in an extra contribution to the kinetic term of $h$, which is not present in the Palatini formulation. Therefore the field transformation to recover a canonically normalised kinetic term is different in the two formulations. Specifically, we have
\bea \label{chi}
  \frac{\rmd\chi}{\rmd h} = \sqrt{\frac{1+\a\xi h^2}{(1+\xi h^2)^2}} \ ,
\eea
where $\chi$ is the canonically normalised field and $\a=1+6\xi$ in the metric formulation and $\a=1$ in the Palatini formulation. From \re{chi} the new field $\chi$ is solved in terms of $h$ as (setting $\chi(h=0)=0$)
\bea \label{chiphi}
  \sqrt{\xi} \chi &=& \sqrt{\a} \,\mathrm{arsinh}(\sqrt{\a\xi}h) - \sqrt{\a-1} \,\mathrm{artanh}\left( \frac{\sqrt{(\a-1)\xi} h}{\sqrt{1+\a\xi h^2}} \right) \ .
\eea
In the Palatini case, the second term is zero, and \re{chiphi} can be inverted as
$\sqrt{\xi}h=\sinh(\sqrt{\xi}\chi)$. In the metric case, there is no general closed form expression for $h$, but in the limit $\xi h^2\gg1$ we have $\sqrt{\xi}h\simeq\frac{1}{2\sqrt{\a}}\left(\frac{1-\b}{1+\b}\right)^{-\ha\b} e^{\sqrt{\frac{\xi}{\a}}\chi}$, where $\b\equiv\sqrt{\frac{\a-1}{\a}}$. This asymptotic form applies in both the metric and the Palatini formulation. In the metric case it reduces to $\sqrt{\xi}h\simeq e^{\sqrt{\frac{1}{6}}\chi}$ for $\xi\gg1$. In this limit, the main difference between the two formulations is that in the metric case we have $\sqrt{1/6}$ in the exponent, and in the Palatini case we have $\sqrt{\xi}\gg\sqrt{1/6}$. We will not use these asymptotic forms, and instead write the potential in terms of the original non-minimally coupled Higgs field $h$, using \re{chi} when taking derivatives with respect to $\chi$.

In the minimally coupled Einstein frame, in terms of the metric $\tilde g$ and scalar field $\chi$, the action \re{actionJ} reads 
\bea \label{actionE}
  S = \int\rmd^4 x \sqrt{-\tilde g} \left( \ha \tilde R - \frac{1}{2} \tilde g^{\a\b} \nabla_\a\chi \nabla_\b\chi - U(\chi) \right) \ ,
\eea
where $\tilde R$ is the Ricci scalar formed from $\tilde g_{\a\b}$ and we have denoted $U(\chi)\equiv V[h(\chi)]\Omega[h(\chi)]^{-4}$.
As the action has the Einstein--Hilbert form, it makes no difference (apart from whether the York--Gibbons--Hawking boundary term has to be included) whether the connection is taken to be an independent variable. The distinction between the metric and Palatini formulations is encoded in the form of the potential $U$ via the field transformation \re{chi}.

At the classical level, the actions \re{actionJ} and \re{actionE} are physically equivalent, as long as physical quantities are properly identified \cite{Chiba:2013, Postma:2014, Jarv:2016sow}. If we consider inflation with a fixed potential, then the predictions for the quantized perturbations are also independent of the frame \cite{Makino:1991, Fakir:1992, Komatsu:1999, Koh:2005, Chiba:2008, Weenink:2010, Kubota:2011}. However, when loop corrections to the effective potential are considered, it is not obvious whether the conformal transformation and quantisation commute.\footnote{Note that there is a continuum of conformal frames, defined by the choice of $\Omega(h)$. Even if the model is defined in the minimally coupled frame to begin with, we could make a conformal transformation to a new classically equivalent frame and quantise there. So this issue is present for any scalar field model of inflation, though it is more explicit in Higgs inflation and other models that are originally formulated in a non-minimally coupled frame.} Various arguments have been presented for frame-independence of quantisation, and no clear evidence for frame-dependence has been found \cite{Weenink:2010, Calmet:2012, Prokopec:2014, Kamenshchik:2014, Burns:2016, Fumagalli:2016, Hamada:2016}. Differences have been found between treatments where renormalisation prescriptions are applied in different frames \cite{Bezrukov:2008b, Bezrukov:2009, Bezrukov:2010, Allison:2013}, but this may be due to not matching the renormalisation scales and integration measures in the different frames consistently \cite{George:2013, Postma:2014, Fumagalli:2016, Hamada:2016}. One possibility to account for this is to use explicitly frame-invariant variables \cite{Prokopec:2012, Postma:2014, Kamenshchik:2014, Prokopec:2014,  Burns:2016}. Also, as pointed out in section 4.2 of \cite{George:2013}, in practice the cosmological perturbations have been calculated by quantising the same field, namely the Sasaki--Mukhanov variable \cite{SM}, just expressed in different frames. It is also argued in \cite{George:2013} that the Einstein frame results should be more reliable than Jordan frame results.\footnote{Note that when quantum effects from the other three modes of the Higgs field are considered \cite{Hertzberg:2010, Mooij:2011, Greenwood:2012, George:2013, George:2015}, it is not possible to simultaneously go to the Einstein frame and get a canonical kinetic term for all of the scalar fields of the Higgs doublet.}

The field redefinition from the Jordan frame to the Einstein frame mixes the scalar field and gravitational degrees of freedom, so if the latter are neglected (as has mostly been the case, with the exception of \cite{Moss:2014, Saltas:2015}), calculations in different frames correspond to different approximations. However, as the dominant loop contributions are from the gauge bosons and the top quark, not from the scalars and gravity, this is not expected to make a large difference. The question of frame equivalence could be settled by doing a full loop calculation, including gravity,  separately in the Jordan and Einstein frames, and comparing the results, although the situation is complicated by the fact that the theory is non-renormalisable.

We simply assume, following \cite{Bezrukov:2014a}, that the loop corrections calculated for the Standard Model in Minkowski spacetime, neglecting gravitons, can be applied in the Einstein frame, with the appropriate choice of field-dependent renormalisation scale. We do not attempt to relate the parameter values that are relevant during inflation with large field values to electroweak scale observables.

\para{Potential and slow-roll parameters.}

In slow-roll inflation, the power spectrum is determined by the shape of the potential. For field values $h\gg v=246$ GeV, the quadratic term can be neglected, so the Jordan frame Higgs potential can be approximated as $V(h)\simeq\frac{1}{4}\l(h)h^4$, with loop corrections being approximated by the field-dependence of $\l(h)$.
The effective potential is gauge-dependent, but the presence of extrema and the field values there are not. As derivatives of the potential are small in slow-roll inflation, gauge-dependence is small, particularly for inflation near an extremum \cite{Cook:2014, Espinosa:2015a, Iacobellis:2016, Espinosa:2016}. Following \cite{Isidori:2007, Hamada:2014, Bezrukov:2014a, Inagaki:2015, Bezrukov:2017}, we use the approximation
\bea \label{l}
  \l(h) = \l_0 + \frac{b}{4} \ln^2\left(\frac{\xi h^2}{\kappa^2 (1+\xi h^2)}\right) \ ,
\eea
which is valid at least near the minimum of $\lambda(h)$, where the logarithm is small (in particular, this holds for $\xi h^2\gtrsim1$ and $\kappa\lesssim1$). We have used the renormalisation prescription I of \cite{Bezrukov:2008b}.

The Einstein frame potential is
\bea \label{U}
  U(\chi) &=& \frac{1}{4} \left[ \l_0 + \frac{b}{4} \ln^2\left(\frac{\xi h^2}{\kappa^2 (1+\xi h^2)}\right) \right] \left( \frac{\xi h^2}{1+\xi h^2} \right)^2 \el
  &\equiv& \frac{\l_0 \kappa^4}{4\xi^2} \left( 1 + c \ln^2 x \right) x^2 \ ,
\eea
where $h(\chi)$ is given by \re{chi} and we have defined $c\equiv\frac{b}{4\l_0}$, $x\equiv\frac{\xi h^2}{\kappa^2(1+\xi h^2)}$ and $\ln^2x\equiv(\ln x)^2$. For the constant $b$ we use the Standard Model value $b=2.3\times10^{-5}$, following \cite{Bezrukov:2014a}. For $\kappa<1$, the parameter $\l_0$ is the value of $\l$ at its minimum, and $\kappa$ sets the location of the minimum. For $\kappa\geq1$, the minimum moves to infinite field values. If we could relate the electroweak scale parameters to the parameters in the inflationary regime, the values of $\l_0$ and $\kappa$ would be fixed; we consider them free parameters. As we will see, the inflationary behaviour can be quite sensitive to the value of $\kappa$. For large values of $\xi h^2$ the coupling approaches the constant $\l_0+b\ln^2\kappa$, and $U$ is asymptotically flat; in terms of the minimally coupled $\chi$ field, it is exponentially flat. We only consider the case when the electroweak vacuum is absolutely stable, and take $\l_0>0$. We consider two possibilities for $\kappa$. The first is $\kappa\sim1$, corresponding to the treatment of \cite{Bezrukov:2014a, Bezrukov:2017}. The second is $\kappa\equiv\kappat\sqrt{\xi}$ with $\kappat\sim1$, the choice used in \cite{Hamada:2014}.

In slow-roll inflation, the power spectrum of curvature perturbations is approximately
\bea
  \mathcal{P}_\sR(k) = A_s \left( \frac{k}{k_*} \right)^{ n_s - 1 + \ha \a_s \ln\frac{k}{k_*} + \frac{1}{6} \b_s \ln^2\frac{k}{k_*}}
\eea
where we use the pivot scale $k_*=0.05$ Mpc$^{-1}$; the amplitude $A_s$, the spectral index $n_s$, the running of the spectrum $\a_s$ and the running of the running $\b_s$ are constant at this level of approximation. They are, to leading order, given by
\bea
  \label{A} A_s &=& \frac{1}{24\pi^2} \frac{U}{\epsilon} \\
  \label{n} n_s &=& 1 - 6 \epsilon + 2 \eta \\
  \label{as} \a_s &=& - 24 \epsilon^2 + 16 \epsilon \eta - 2 \sigma_2 \\
  \label{bs} \b_s &=& - 192 \epsilon^3 + 192 \epsilon^2 \eta - 32 \epsilon \eta^2 - 24 \epsilon \sigma_2 + 2 \eta \sigma_2  + 2 \sigma_3 \ ,
\eea
and the tensor-to-scalar ratio $r$ is
\bea
  r = 16 \epsilon \ ,
\eea
where the slow-roll parameters are
\bea \label{SR}
  \epsilon &=& \ha \left( \frac{U'}{U} \right)^2 \ , \qquad \eta = \frac{U''}{U} \el
  \sigma_2 &=& \frac{U'}{U} \frac{U'''}{U} \ , \qquad\quad \sigma_3 = \left( \frac{U'}{U} \right)^2 \frac{U''''}{U} \ ,
\eea
where prime denotes derivative with respect to $\chi$; when calculating the slow-roll parameters from \re{U}, we use \re{chi} to convert to derivatives with respect to $h$. The lowest order slow-roll parameters and the amplitude are
\bea
  \label{epsilon} \epsilon &=& \frac{8}{(1+\a\xi h^2) h^2} \left( \frac{1 + c\ln x + c\ln^2 x}{1 + c\ln^2 x} \right)^2 \el
  \label{eta} \eta &=& \frac{4}{(1+\a\xi h^2)^2 h^2} \frac{1}{1+c\ln^2 x} \left( 3 + [2\a-1] \xi h^2 - 2 \a\xi^2h^4 + 2 [1+\a\xi h^2] c \right. \el
  && \left. + [ 7 + (6\a-1)\xi h^2 - 2\a\xi^2h^4 ] c \ln x + [ 3 + (2\a-1)\xi h^2 - 2\a\xi^2h^4 ] c \ln^2 x \right) \el
  \label{Apot} \frac{U}{\epsilon} &=& \frac{\l(h)}{32} h^6 \frac{1+\a\xi h^2}{(1+\xi h^2)^2} \left( \frac{1 + c\ln^2 x}{1 + c \ln x + c\ln^2 x} \right)^2 \ ,
\eea
with $\l(h)=\l_0 (1 + c\ln^2 x)$. In slow-roll, $\epsilon<1$ and $|\eta|<1$. Also, from \re{n} and the observational value $n_s=0.9569\pm0.0077$ \cite{Planck:inflation}\footnote{This value of $n_s$ is based on the Planck TT+lowP dataset, assumes the \LCDM model and includes running and running of the running. The central value and error bars of $n_s$ depend slightly on which datasets and parameters are considered \cite{Planck:inflation, Cabass:2016}. The spectrum is close to featureless and almost scale-invariant, with a red tilt, even in a parameter-free reconstruction, still assuming the \LCDM model \cite{Ravenni:2016}.} we know that $3\epsilon_*-\eta_*=(2.155\pm0.39)\times10^{-2}\approx2\times10^{-2}$, where star denotes the value at the time when the pivot scale exits the Hubble radius. The observational upper limit $r_*<0.07$ (assuming no running of $n_s$) \cite{r} implies $\epsilon_*<5\times10^{-3}$. (We quote error bars as 68\% confidence intervals and limits as 95\% confidence intervals.)

\begin{figure}[t]
\center
\includegraphics[width=.75\textwidth]{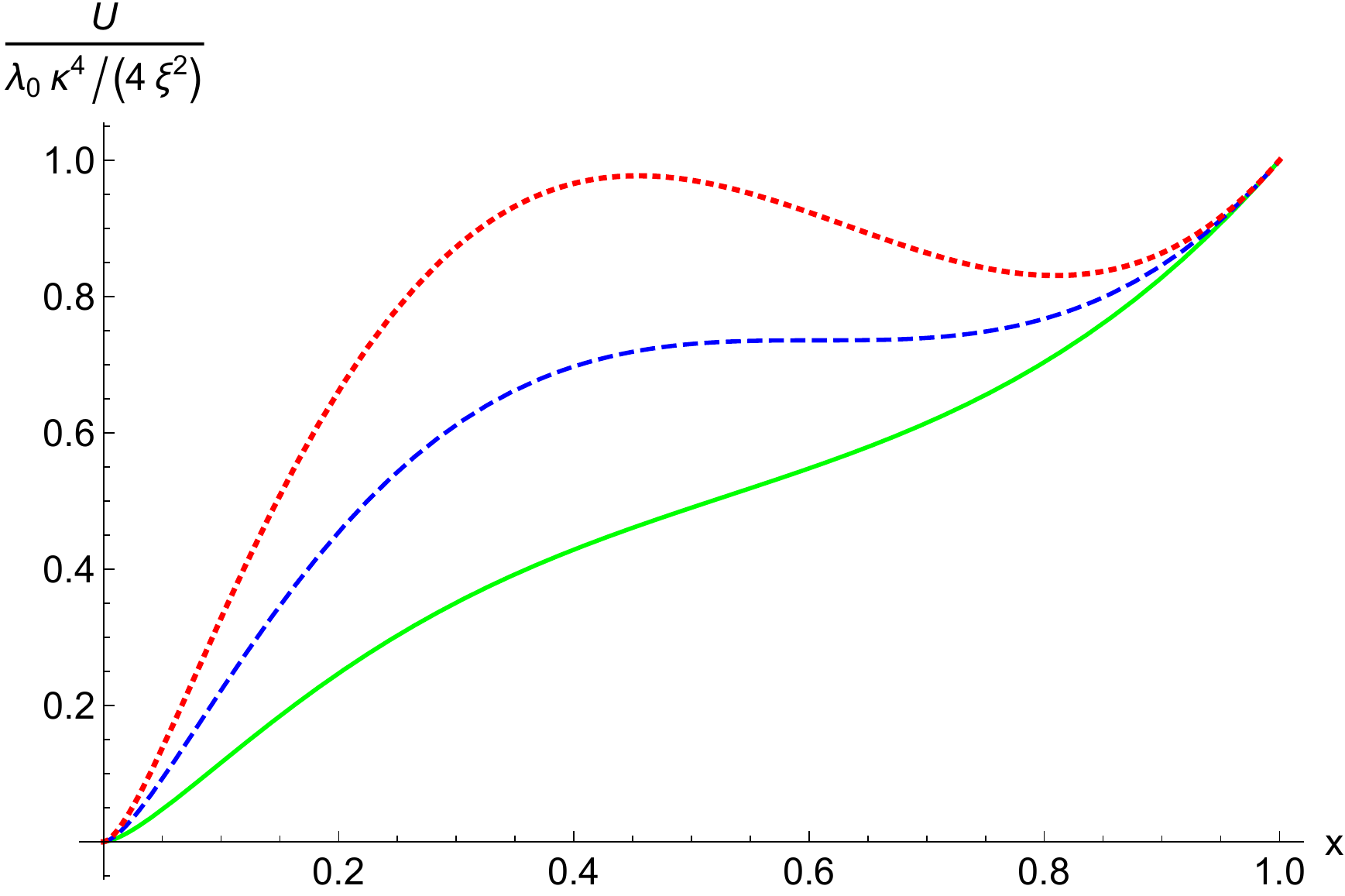}\\
\caption{The potential as a function of $x$ for $c=2,4,6$ (solid green, dashed blue, dotted red). The top curve is relevant for hilltop and false vacuum inflation, middle curve for critical point inflation and the bottom curve for plateau inflation.}
\label{fig:pot}
\end{figure}

If $c<4$, the only minimum of the potential is at $h=0$. At $c=4$, corresponding to $\l_0=\frac{b}{16}=1.4\times10^{-6}$, there is an inflection point (with $U''=0$, and also $U'=0$). For $c>4$, a new minimum appears at large field values, and there is a hilltop between this false vacuum and the electroweak vacuum. (The inflection point requires $\kappa<e^{1/4}$, while the hilltop can exist for $\kappa<e^{1/4}\ldots e^{1/2}$, depending on the value of $c$.)
In \fig{fig:pot} we show the form of the potential for different values of $c$. Note that the potential is plotted as function of $x$, so its curvature cannot be readily estimated by eye, requiring conversion from $x$ to $h$, and from $h$ to $\chi$ via \re{chi}. For example, the short upturn at values $x\approx1$ corresponds to an exponentially flat plateau extending to infinite values in terms of $\chi$.

In addition to the plateau, inflation is possible at the critical point, at the hilltop and in the false vacuum. The approximation \re{l} for the coupling is optimised to be accurate near the minimum of $\l$, and may not be accurate at the hilltop. Nevertheless, we consider also the hilltop case, to get an understanding of the differences between the metric and the Palatini formulations, as the potential \re{U} has qualitatively the right shape.
In the false vacuum case, non-minimal coupling is not needed to sustain inflation, and our approximation for the potential is not valid. However, new physics is required for graceful exit, which may involve the non-minimal coupling of another scalar field to gravity \cite{Masina:2011a}. We will consider the difference the Palatini formulation makes for the dynamics of that field. 

There are two ways for $\epsilon$ to be small. Either $(1+\a\xi h^2)h^2\gg1$ or $|1 + c \ln x + c\ln^2 x|\ll1$. The first possibility corresponds to plateau inflation, the second to critical point inflation and hilltop inflation (in false vacuum inflation the potential has a qualitatively similar shape as in hilltop inflation). Let us consider them in turn.

\subsection{Plateau inflation} \label{sec:plateau}

\para{Metric formulation.}

Let us first consider the case $(1+\a\xi h^2)h^2\gg1$ in the metric formulation. We assume $6\xi\gg1$; if this condition is not satisfied, the results of the metric formulation agree with those of the Palatini formulation, which we consider next. The condition $(1+\a\xi h^2)h^2\gg1$ then reduces to $6\xi^2h^4\gg1$, and we have
\bea
  \epsilon &\simeq& \frac{4}{3 (\xi h^2)^2} \left( \frac{1 + c\ln x + c\ln^2 x}{1+c\ln^2 x} \right)^2 \simeq \frac{3}{4} \eta^2 \el
  \eta &\simeq& - \frac{4}{3 \xi h^2} \frac{1 + c\ln x + c\ln^2 x}{1 + c\ln^2 x} \el
  \frac{U}{\epsilon} &\simeq& \frac{3 \l(h)}{16 \xi^2} (\xi h^2)^2 \left( \frac{1 + c\ln^2 x}{1 + c \ln x + c\ln^2 x} \right)^2 \simeq \frac{\l(h)}{3 \xi^2} \eta^{-2} \ .
\eea
The condition $3\epsilon_*-\eta_*=0.02$ gives (taking into account $\epsilon_*<5\times10^{-3}$) $\epsilon_*=3\times10^{-4}$ and $\eta_*=-0.02$, so we have $r=5\times10^{-3}$. The observed amplitude $U/\epsilon=24\pi^2 A_s=5\times10^{-7}$ \cite{Planck:inflation} is reproduced for $\l_*/\xi^2=6\times10^{-10}$. If the running of the quartic coupling from the measured electroweak scale value $\l(100\ \mathrm{GeV})=m^2/(2 v^2)=0.13$ is neglected, this implies $\xi=1\times10^4$. These are the results of the original Higgs inflation proposal \cite{Bezrukov:2007}. If loop corrections are taken into account, $\xi$ can in principle be made as small as desired by adjusting $\l_*=\l_0 (1 + c\ln^2 x_*)\simeq\l_0 (1 + 4 c\ln^2\kappa)$ to be small. If we want to avoid having a false vacuum, we must have $c\leq4$ and thus $\l_*\sim\l_0\gtrsim1.4\times10^{-6}$, giving $\xi\gtrsim50$. For $\kappa\leq1$, the false vacuum is at smaller field values than the inflationary plateau, preventing the field from slowly rolling to the electroweak minimum. However, with $\kappa>1$, it is possible to push the false vacuum beyond the inflationary plateau to avoid this constraint; see the discussion of critical point inflation in \sec{sec:crit}.

The non-minimal coupling $\xi$ is a free parameter, but $\eta$ depends on the number of e-folds $N$. Therefore if we calculate $N_*$ independently, the value of $\eta_*$ provides a consistency check. The relation between $\xi h^2$ and $N$ is given by
\bea \label{N}
  N &=& \int^{\chi}_{\chi_\rme} \rmd\chi \frac{U}{U'} = \frac{1}{4} \int^{h}_{h_\rme} \rmd h \frac{1+\a\xi h^2}{1+\xi h^2} h \frac{1+c\ln^2 x}{1 + c\ln x + c\ln^2 x} \el
  &\approx& \frac{\a}{8} ( h^2 - h_\rme^2 ) - \frac{3}{4} \ln\frac{1+\xi h^2}{1+\xi h^2_\rme} \el
  &\approx& \frac{3}{4} \xi h^2 \ ,
\eea
where end refers to the end of inflation, on the first line we have used \re{chi} to change the integration variable to $h$, on the second line we have assumed that the loop contribution can be dropped and on the third line applied $6\xi\gtrsim1$, $\xi h^2\gg\xi h_\rme^2$ and $\xi h^2\gg1$. Note that if $4-c\ll1$ so that there is a critical point, a significant number of the total e-folds can be accumulated there even if inflation happens on the plateau, changing this mapping between $N$ and $h$.

The number of e-folds at the pivot scale $k_*=0.05$ Mpc$^{-1}$ is
\bea \label{Npivot}
  N_* &=& 61 - \Delta N_\reh + \frac{1}{4} \ln\frac{U_*}{U_\rme} + \frac{1}{4}\ln U_* \el
  &=& 52 + \frac{1}{4} \ln\frac{r}{0.07} \ ,
\eea
where $\Delta N_\reh$ is the number of e-folds between the end of inflation and the end of preheating (defined as the time when energy density starts to scale like radiation). On the second line we have taken into account that for Standard Model field content $\Delta N_\reh=4$ \cite{Figueroa:2009,Figueroa:2015,Repond:2016} (although see \cite{Ema:2016}), written $U=\frac{3\pi^2}{2} A_s r$, inputted the observed value $24\pi^2 A_s=5\times10^{-7}$, inserted the maximum value $r_*=0.07$ allowed by observations as a point of comparison, and dropped the term due to the change of the potential between the pivot scale and the end of inflation (it gives a correction that is $<1$).
The pivot scale thus corresponds to $N_*=50$ (or $\xi h_*^2=70$), so we get $\eta_*=-\frac{1}{N}=-0.02$, $\epsilon_*=\frac{3}{4 N^2}=3\times10^{-4}$. The spectral index $n_s=1-\frac{2}{N}-\frac{9}{2N^2} =0.96$ and the tensor-to-scalar ratio $r=\frac{12}{N^2}=5\times10^{-3}$ are thus predictions without free parameters.

As $\xi h_*^2$ is fixed, making $\xi$ smaller correspondingly increases $h_*$, which in principle imposes stronger constraints on the observed top and Higgs masses to keep the potential positive; however, this assumes that we would know how to run the electroweak scale parameters to the inflationary scale. For $\xi\lesssim70$, the field value will be larger than the Planck scale, which, as discussed in \sec{sec:pot}, is not necessarily problematic. (The value of the minimally coupled field $\chi$ becomes trans-Planckian as well.) If $\xi$ is small, $6\xi\ll1$, the above approximations do not hold. However, in that case, the metric and Palatini results agree, so it is included in our discussion of the Palatini formulation.

\para{Palatini formulation.}

In the Palatini case, discussed in \cite{Bauer:2008}, the e-fold relation \re{N} gives $N_*=\frac{1}{8}h_*^2$ (taking into account $h\gg h_\rme$), so $N_*=50$ corresponds to $h_*=20$, and
\bea \label{platpala}
  \epsilon_* &=& 2\times10^{-2} \frac{1}{1+\xi h_*^2} \left( \frac{1 + c\ln x_* + c\ln^2 x_*}{1+c\ln^2 x_*} \right)^2 \el
    \eta_* &=& 10^{-2} \frac{1}{(1+\xi h_*^2)^2} \frac{1}{1+c\ln^2 x_*} \left( 3 + \xi h_*^2 - 2 \xi^2h_*^4 + 2 [1+\xi h_*^2] c \right. \el
  && \left. + [ 7 + 5 \xi h_*^2 - 2 \xi^2h_*^4 ] c \ln x_* + [ 3 + \xi h_*^2 - 2\xi^2h_*^4 ] c \ln^2 x_* \right) \el
  \frac{U_*}{\epsilon_*} &=& 2\times10^{6} \l_* \frac{1}{1+\xi h_*^2} \left( \frac{1 + c\ln^2 x_*}{1 + c \ln x_* + c\ln^2 x_*} \right)^2 \ .
\eea

Without the loop correction, we would have $r_*=0.32/(1+\xi h_*^2)$, the result of $\l h^4$ inflation suppressed by the factor $1+\xi h_*^2$. Getting the scalar power spectrum amplitude right requires $\l_*/(1+\xi h_*^2)\sim10^{-13}$. While the quartic coupling can be arbitrarily small at high field values due to loop corrections (as it can reach zero), its dependence on the field value spoils the flatness of the potential. Also, the potential must be such that the field will end up in the electroweak vacuum from the inflationary region. Because of this, successful inflation is not possible if $\xi h_*^2\ll1$ and the non-minimal coupling does not contribute \cite{Isidori:2007, Hamada:2013, Fairbairn:2014}. Because we then have $\xi\ll1$, the metric and Palatini formulations behave in the same way. If $\xi h_*^2\sim1$ there can be cancellations in the numerator of $\eta$, so we consider this case only numerically. In the limit $\xi h_*^2\gg1$, we have
\bea
  \epsilon_* &=& 5 \times 10^{-5} \frac{1}{\xi} \left( \frac{1 + c\ln x_* + c\ln^2 x_*}{1+c\ln^2 x_*} \right)^2 \el
  \eta_* &=& - 2\times10^{-2} \frac{1 + c\ln x_* + c\ln^2 x_*}{1+c\ln^2 x_*} \el
  \frac{U_*}{\epsilon_*} &=& 5 \times10^3 \frac{\l_*}{\xi} \left( \frac{1 + c\ln^2 x_*}{1 + c \ln x_* + c\ln^2 x_*}\right)^2 \ .
\eea
The spectral index condition $3\epsilon_*-\eta_*=0.02$ gives $c\ln x_*=-8\times10^{-3}/\xi\ll1$. The running at the inflationary scale is thus small, and the normalisation condition gives $\l_*/\xi=10^{-10}$, as found in \cite{Bauer:2008}. We thus require an even larger $\xi$ than in the metric formulation for the same $\l_*$. For $\l_*=0.1$ we would have $\xi=10^9$ and $r=10^{-12}$. (In this case, the second term in the number of e-folds \re{N} is $-6$, so $N=46$; as in the metric formulation, for $4-c\ll1$ there can be a contribution to $N$ from the critical point.) The condition $\l_0>10^{-6}$ to avoid a false vacuum gives $\xi>10^4$. The tensor-to-scalar ratio is $r=8\times10^{-4}/\xi=8\times10^{-14}/\l_*\lesssim10^{-7}$. In terms of $N$ we have $\eta_*=-\frac{1}{N}=-0.02$ and $\epsilon_*=\frac{1}{8\xi N^2}\ll|\eta_*|$. So, to leading order in $N$, $n_s$ is the same as in the metric case, but $r$ is suppressed by $6\xi$, $r=\frac{2}{\xi N^2}$.

Before reporting on the numerical study of the full parameter space in \sec{sec:num}, let us consider inflation away from the plateau.

\subsection{Critical point inflation} \label{sec:crit}

\para{Critical point condition.}

We now look at the possibility that $\epsilon$ is small not because $(1+\a\xi h^2)h^2$ is large, but because loop corrections are significant, so that \mbox{$|1+c\ln x+c\ln^2 x|\ll1$}. This is possible only if $c\gtrsim4$. In the case when $1+c\ln x+c\ln^2 x=0$ (possible for $c\geq4$), the potential \re{U} has another extremum (in addition to the one at $h=0$). The second slow-roll parameter \re{eta} shows that there is an inflection point at $\ln x=-\ha$ if and only if $c=4$ \cite{Isidori:2007}. This case is illustrated in \fig{fig:pot} with a dashed blue line. This means that $\xi h^2=e^{-1/2}\kappa^2/(1-e^{-1/2}\kappa^2)\equiv\xi h_c^2$ and $\l=2\l_0=b/8=2.9\times10^{-6}\equiv\l_c$. Let us first consider the case $\kappa\sim1$. For $\kappa=1$ we have $\xi h^2=1.5$, whereas for $\kappa\rightarrow e^{1/4} \approx 1.3$.
the inflection point is pushed to infinitely large values of $\xi h^2$. This means that in the case $\kappa=\kappat\sqrt{\xi}$, we must have $\xi<e^{1/2}\kappat^{-2}$, so the demand that there is an inflection point restricts $\xi$ to small values.

Inflation near an inflection point was first studied in the context of Minimal Supersymmetric Standard Model \cite{Allahverdi:2006a, Allahverdi:2006b}, and subsequently also in the context of a string-inspired model \cite{BuenoSanchez:2006}. Inflection point Higgs inflation was also recently studied in the context of a $B-L$ extended Standard Model \cite{Okada:2016}.
The possibility of Higgs inflation at the critical point was first considered in \cite{Allison:2013}. It was studied in detail in \cite{Bezrukov:2014a, Hamada:2014, Bezrukov:2014b, Rubio:2015} (see also \cite{Fumagalli:2016, Enckell:2016, Bezrukov:2017}) to accommodate the claimed detection of inflationary gravitational waves with $r\approx0.2$ by the BICEP2 telescope (the signal was since shown to be due to foregrounds \cite{fore}).

As the loop corrections are important, the potential is a bit complicated, and the numerical results in \sec{sec:num} are essential for fully investigating the parameter space. However, to get some understanding of the results, let us look analytically at the lowest order slow-roll parameters close to the critical point. Let us fix $x=e^{-\ha}\equiv x_c$ and write $c\equiv4(1-\d_c)$. The first two slow-roll parameters and the amplitude are
\bea \label{critpala}
  \epsilon &=& \frac{2}{(1+\a\xi h_c^2) h_c^2} \frac{\d_c^2}{(1-\ha\d_c)^2} \\
  \eta &=& \frac{2}{(1+\a\xi h_c^2)^2 h_c^2} \left( 3 + [2\a-1] \xi h_c^2 - 2 \a \xi^2h_c^4 \right) \frac{\d_c}{1-\ha\d_c} \\
  \frac{U}{\epsilon} &=& \frac{\l_c}{8} h_c^6 \frac{1+\a\xi h_c^2}{(1+\xi h_c^2)^2} \frac{(1-\ha\d_c)^2}{\d_c^2} \ .
\eea

\para{Metric formulation.}

Let us first consider the metric case that has been studied in the literature. As before, we take $6\xi\gtrsim1$; if this condition is not satisfied, the results agree with the Palatini formulation. Let us first consider $\kappa\sim1$. If we take $\kappa=1$, we have $\xi h_c^2=1.5$, and further assuming $\d_c\ll1$ we have (assuming that $\xi\gtrsim1$, otherwise there can be a cancellation in the numerator of $\eta$)
\bea \label{critmetric}
  \epsilon_* &=& 0.14 \d_c^2 \\
  \eta_* &=& - 0.15 \d_c \\
  \frac{U_*}{\epsilon_*} &=& 2 \times 10^{-6} \frac{1}{\xi^2\d_c^{2}} \ .
\eea
From the spectral index condition we get $\d_c=0.1$, and $U_*/\epsilon_*=5\times10^{-7}$ then gives $\xi=20$. The tensor-to-scalar ratio is $r_*=0.02$.

If we instead tune $\kappa$ so that $\xi h_c^2\gg1$, and keep $\d_c\ll1$, we have
\bea 
  \epsilon_* &=& \frac{1}{3} \frac{\d_c^2}{\xi^2 h_c^4} \\
  \eta_* &=& - \frac{2}{3} \frac{\d_c}{\xi h_c^2} \\
  \frac{U_*}{\epsilon_*} &=& 2 \times 10^{-6} \frac{\xi^2 h_c^4}{\d_c^2} \frac{1}{\xi^2}\ .
\eea
Now the spectral index condition gives $\xi h_c^2/\d_c=35$, so $r=4\times10^{-3}$, and the correct amplitude is reproduced for $\xi=70$. (Note that there is some tension between the assumptions $\xi h_c^2=35\d_c\gg1$ and $\d_c\ll1$.)

In the case $\kappa=\kappat\sqrt{\xi}$, we must have $\kappat=1/\sqrt{20}\approx0.2$ or $\kappat=e^{1/4}/\sqrt{70}\approx0.2$ to get the same results as above.

We see that $r$ can be much larger than in the plateau inflation case and its value depends sensitively on $\kappa$. However, we also have to check that the number of e-folds is correct, and also take into account that the field may not be exactly at the critical point. We will do this numerically, reproducing the results of \cite{Hamada:2014, Bezrukov:2014a}.
Before presenting the numerical plots, let us look at critical point inflation analytically in the Palatini case, which has not been considered before.

\para{Palatini formulation.}

Taking first $\kappa=1$ and $\d_c\ll1$, we have
\bea
  \epsilon &=& 0.51 \xi \d_c^2 \\
  \eta &=& - 0.04 \xi \d_c \\
  \frac{U}{\epsilon} &=& 5 \times 10^{-7} \frac{1}{\xi^3\d_c^2} \ .
\eea
Now the spectral index and amplitude conditions give $\d_c=0.02$ and $\xi=14$, leading to $r_*=0.05$. It seems possible, as in the metric formulation, to have $r$ at the level of the current observational limit, though we have to check that the number of e-folds comes out correct.

If we instead tune $\kappa$ to get $\xi h_c^2\gg1$, we have
\bea
  \epsilon &=& 2 \frac{\d_c^2}{h_c^4} \frac{1}{\xi} \\
  \eta &=&  - 4 \frac{\d_c}{h_c^2} \\
  \frac{U}{\epsilon} &=& 4 \times 10^{-7} \frac{h_c^4}{\d_c^2} \frac{1}{\xi} \ ,
\eea
from which we get $h_c^2/\d_c=160$ and $\xi=2\times10^4$, which give $r_*=6\times10^{-8}$.

In the case $\kappa=\kappat\sqrt{\xi}$, the first example implies $\kappat=1/\sqrt{14}\approx0.3$, while the second would require a very small $\kappat=e^{1/4}/\sqrt{2\times10^4}\approx9\times10^{-3}$. In any case, the results have to be checked by numerical calculation, which we present in \sec{sec:num}.

\para{Slow-roll hierarchy in the Palatini formulation.}

In plateau inflation, in the metric formulation we have $U_*/\epsilon_*\sim\l_*/(\eta_*^2\xi^2)\sim10^{-6}$. Because $\eta_*\sim10^{-2}$, we have $\l_*/\xi^2\sim10^{-10}$. In the Palatini formulation we instead get $\l_*/\xi\sim10^{-10}$. In critical point inflation, we have roughly the same result for $\xi h_c^2\gg1$, and because $\l_*\sim\l_c\sim10^{-6}$ is fixed, in the metric case we get $\xi\sim\eta_*^{-1}\sim100$ and in the Palatini case $\xi\sim\eta_*^{-2}\sim10^4$. (For $\xi h_c^2\sim1$, cancellations in the numerator of $\eta$ can change these order of magnitude estimates, as we have seen.) This difference between the metric and Palatini formulations is related to the fact that for $\xi h^2\gg1$, each derivative of the potential with respect to $\chi$ brings down a factor of $\sqrt{1/6}$ in the metric case and $\sqrt{\xi}$ in the Palatini case.

The fact that every derivative with respect to $\chi$ brings down a factor of $\sqrt{\xi}$ in the Palatini case may raise concerns about whether the slow-roll hierarchy holds when $\xi\gg1$. However, in the $l$:th order slow-roll parameter the $l+1$:th order derivative of $H(\chi)$ is always accompanied by the factor $\frac{1}{(2 H)^{l-1}}\frac{\rmd^{l-1} H}{\rmd \chi^{l-1}}=\epsilon^{(l-1)/2}$ \cite{Liddle:1994}. (The argument is formulated in terms of derivatives of the Hubble parameter $H$, but it works similarly for $U$, as in our slow-roll parameters $\sigma_2$ and $\sigma_3$ in \re{SR}.) Therefore the slow-roll hierarchy is not disturbed if $\sqrt{\xi\epsilon}\ll1$. From \re{platpala} and \re{critpala} we see that this condition is satisfied (assuming the relation between $h_*$ and $N$ is not drastically altered by the loop corrections).
However, as the Palatini first order slow-roll parameters are amplified by $\xi$ compared to the metric formulation, the field has to be nearer the critical point to reach the same slow-roll parameters, as we have seen. Similarly, the running of the spectral index is enhanced in the Palatini case.

\subsection{Hilltop inflation} \label{sec:hill}

\para{Hilltop condition.}

Let us now consider hilltop inflation \cite{Boubekeur:2005, Barenboim:2016}. For $c>4$ the potential develops a false vacuum, as depicted in \cref{fig:pot} with a red dotted line. In this case inflation can happen near the top of the hill \cite{Fumagalli:2016}. This setup is sensitive to initial conditions, because the field has to start close to the top of the hill and roll down to the side of the electroweak vacuum after inflation. We have used an approximation for $\l(h)$ that is optimised around a minimum of $\l(h)$, which may not be valid close to the hilltop. The shape of the potential is nevertheless qualitatively correct, so we use it to illustrate the differences between the metric and Palatini formulations, a detailed analysis of loop corrections at the hilltop being outside the scope of this paper.

Assuming $c>4$, the potential \eqref{U} has a maximum at $x=\xh$, with
\bea
  \ln \xh = - \half - \half \sqrt{ 1 - 4/c } \ .
\eea
This condition implies that we must have $\kappa<e^{1/4}\ldots e^{1/2}$, depending on the value of $c$.
Writing $x = \xh (1-\deh)$ and assuming that the field is on the side of the electroweak vacuum near the maximum, $0<\deh\ll1$, we get
\bea
  \epsilon &=& \frac{32}{ (1 + \a\xi h^2) h^2 } \frac{ 1 - 4/c }{ ( 1 + \sqrt{1-4/c} )^2 } \deh^2 \\
  \eta &=& - \frac{16}{ (1 + \a\xi h^2) h^2 } \frac{ \sqrt{1 - 4/c} }{ 1 + \sqrt{1-4/c} } \\
  \frac{U}{\epsilon} &=& \frac{b}{1024} h^6 \frac{1+\a\xi h^2}{(1 + \xi h^2)^2} \frac{ (1 + \sqrt{1-4/c})^3 }{ 1 - 4/c } \deh^{-2} \ .
\eea
The tensor-to-scalar ratio is suppressed, but being near the hilltop (unlike being near the critical point) does not automatically guarantee $|\eta|\ll1$. For that, we need either $c-4\ll1$ (\ie the hilltop has to be close to being an inflection point) or $\xi h^2\gg1$. In either case, we have an another small parameter in addition to $\deh$ (also, $\xi h^2$ and $h^2$ occur separately), so the spectral index and the amplitude are not enough to determine the tensor-to-scalar ratio. The results are also sensitive to getting the right number of e-folds, which has to be determined numerically, so we will not discuss the analytical approximation in the hilltop case further.

\subsection{False vacuum inflation} \label{sec:false}

\para{Graceful exit and non-minimal coupling.}

Let us now consider false vacuum Higgs inflation \cite{Masina:2011a, Masina:2011b, Masina:2012, Masina:2014, Notari:2014, Fairbairn:2014, Iacobellis:2016}. The qualitative shape of the potential is again illustrated in \fig{fig:pot} with a dotted red line. During inflation the Higgs field rests in the second minimum and dominates the energy density. Note that non-minimal coupling of the Higgs field is not needed to get a false vacuum, it can be achieved with loop corrections in the Standard Model without gravity. However, physics beyond the Standard Model is needed to lower the potential barrier to allow for a graceful exit. Possible mechanisms include an extra scalar field, which does not directly couple to the Higgs field \cite{Masina:2011a} (similar to the model studied in \cite{Biswas:2005} in a non-Higgs context), and a hybrid inflation scenario \cite{Masina:2012, Fairbairn:2014}. In the former case, non-minimal coupling of the extra field plays an important role. In \cite{Notari:2014} a non-minimal coupling was also added for the Higgs field to make the minimum more shallow. However, there the potential is written in terms of a Higgs field whose kinetic term has not been transformed to the canonical form, in which case there is no difference between the metric and Palatini formulations. To compare the two, the effects of this transformation on the analysis in \cite{Notari:2014} would have to be worked out. This is outside the scope of the present paper, and we will just consider the model of \cite{Masina:2011a}.

The probability for the Higgs field to tunnel from the false vacuum grows with decreasing $H$. In \cite{Masina:2011a}, an extra field non-minimally coupled to gravity is used to decrease $H$ and trigger the tunnelling. The action is
\bea \label{actionJfalse}
  S &=& \int\rmd^4 x \sqrt{-g} \left( \frac{1}{2} f(\phi) g^{\a\b} R_{\a\b}(\Gamma,\pat\Gamma) - \frac{1}{2} g^{\a\b} \nabla_\a\phi \nabla_\b\phi \right. \el
  && \left. - \frac{1}{2} g^{\a\b} \nabla_\a h \nabla_\b h - V(h) \right) \ ,
\eea
where $h$ is the Higgs field as before, $\phi$ is the extra field and $f$ is a positive function; note that the Higgs is minimally coupled. As discussed in \cref{sec:pot}, the action can be brought to the minimally coupled form by the conformal transformation
\bea
  g_{\a\b}=\Omega^{-2}\tilde g_{\a\b}=f^{-1}\tilde g_{\a\b} \ .
\eea
The canonical kinetic term is restored by transforming the field as
\bea \label{falseTrans}
  \frac{\rmd\Phi}{\rmd\phi} = \sqrt{\frac{2 f + \frac{\a-1}{2\xi} f'^2}{2 f^2}} \ ,
\eea
where $\Phi$ has a canonical kinetic term and prime denotes differentiation with respect to the auxiliary field $\phi$. The term $\frac{\a-1}{2\xi}$ is 3 in the metric case and 0 in the Palatini case. The Higgs potential becomes
\bea \label{falsepot}
  U(h, \Phi) = \frac{V(h)}{f[\phi(\Phi)]^2} \ .
\eea

In \cite{Masina:2011a} it is assumed that inflation begins at small values of $\phi$, where the observed part of the spectrum is also generated, and ends at large values. It is assumed that the function $f$ can be expanded as a power series
\bea \label{seriesF}
  f(\varphi) = 1 + \sum_n \c_n \phi^n
\eea
where $\c_n$ are constants. When $\phi\ll1$, the field transformation agrees to leading order in the metric and Palatini formulations, but when the field is large, the situation is different. In the metric case, if the function $f$ grows faster than $\phi^2$, the field transformation becomes approximately
\bea
  \frac{\rmd\Phi}{\rmd\phi} = \sqrt{\frac{3}{2}} \frac{f'}{f} \ ,
\eea
so the Einstein frame potential is completely independent of the specific form of $f$ (note that $h$ is constant during inflation),
\bea
  U(h, \Phi) = V(h) e^{- 2 \sqrt{\frac{2}{3}} \Phi} \ .
\eea
As $\Phi$ rolls to larger field values, $U$ and therefore $H$ decrease, and the Higgs can tunnel out, ending inflation. In the Palatini case, the field transformation does not involve $f'$ and depends on the form of $f$. Taking $f(\phi) = 1 + \xi \phi^2$ gives
\bea
  U(h, \Phi) = \frac{V(h)}{\cosh^4(\sqrt{\xi} \Phi)} \simeq 16 V(h) e^{-4 \sqrt{\xi} \Phi} \ ,
\eea
Unlike in the metric case, it is possible to adjust the steepness of the potential, but this has no effect on the small-field stage, only the end of inflation. Although the details are different from the metric case, the overall picture is the same: as $\Phi$ grows, $H$ decreases and the Higgs eventually tunnels out. Any coupling of the form $f(\phi) = 1 + \c_n \phi^n$ leads to qualitatively the same kind of evolution. As the potential \re{falsepot} is the same in both cases, the slow-roll parameters are the same in terms of the number of e-folds, but the number of e-folds can be different due to different evolution in the late stages of inflation and after it.

\section{Numerical results} \label{sec:num}

\subsection{Setup}

\para{Critical point inflation.}

In order to investigate the entire parameter space we use the slow-roll equations without any further assumptions on the parameters, and calculate the amplitude, spectral index and its running, and running of the running, numerically. Overall we have five parameters: $\xi$, $\kappa$, $\lambda_0$, $h_*$ and $h_\rme$. We have three constraint equations: the normalisation condition \eqref{A}, the constraint on the number of e-folds \eqref{Npivot} and the condition $\epsilon=1$ or $|\eta|=1$ at the end of inflation. This leaves two free parameters, which we choose to be $\xi$ and $\kappa$, from which we convert to the $(n_s,r)$ plane for comparison with observations and previous studies. For $\kappa$, we consider two possibilities. Either we take $0.1<\kappa<10$, or $\kappa=\sqrt{\xi}\kappat$ with $0.1<\kappat<10$, corresponding to the choices in \cite{Bezrukov:2014a, Bezrukov:2017} and \cite{Hamada:2014}, respectively. 
Since this parameter range includes inflation near an almost inflection point, which we are particularly interested in, we put it under the label of critical point inflation, although it also includes inflation far from the critical point.

We determine $h_*$ from the normalisation condition \eqref{A}. For some regions of the parameter space the normalisation condition can be degenerate with up to three solutions. Usually we have only one solution, but when $\lambda_0$ is near the critical value, then $\epsilon$ is close to zero near the critical point. As the amplitude is inversely proportional to $\epsilon$, it has a peak, which can lead to additional solutions. The derivative of the amplitude with respect to the field is
\bea
  \frac{\rmd A_s}{\rmd\chi} = (1-n_s) \frac{A_s}{\sqrt{2 \epsilon}} \ ,
\eea
so the spectrum is blue (red), when the derivative of the amplitude is negative (positive). The derivative must change its sign between solutions, because we are comparing the theoretical value with the measured constant value, and this leads to alternating red and blue solutions.
We have checked numerically that the smallest field value does not give enough e-folds and is hence excluded. The middle one gives a blue spectrum, and is excluded by observations. The largest field value is the only one to meet the observational criteria and we choose it whenever the normalisation condition is degenerate.

We determine $h_\rme$ from the condition $\epsilon=1$ or $|\eta|=1$, whichever is reached first. As with the normalisation condition, the end condition can have up to three solutions, so there can be disjoint slow-roll regions with fast-roll regions in between. In particular, there can be a fast-roll region between the inflationary plateau and the critical point. We choose the smallest field value to be the end point of inflation and disregard any fast-roll regions between the pivot scale and the end of inflation (the effect of this on the number of e-folds is minimal).

Finally, we determine $\lambda_0$ by demanding that the number of e-folds after the pivot scale leaves the Hubble radius agrees with the constraint \eqref{Npivot}, with $\Delta N_\text{reh}=4$, following \cite{Figueroa:2009,Figueroa:2015,Repond:2016}. Taking $r$ into account, this leads to between 45 and 51 e-folds. We have also studied the effect of changing the duration of reheating, and comment on how the results would change.

\para{Hilltop inflation.}

For hilltop inflation the field has to start below the hilltop on the side of the electroweak vacuum, and we must have $c>4$. This guarantees a unique solution for both the normalisation condition and the end of inflation condition. As in critical point inflation, we use the number of e-folds to determine $\lambda_0$. We use the same limits for $\kappa$ and the same range of e-folds as in the critical point case. 

\subsection{Critical point inflation}

\para{Metric formulation.}

\begin{figure}[t]
\center
\begin{minipage}[t]{.45\textwidth}
\includegraphics[width=\textwidth]{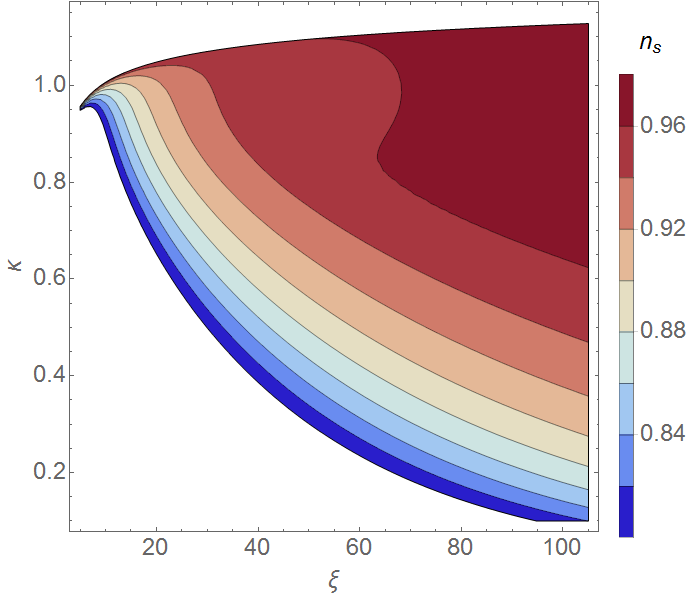}
\caption{Spectral index in the metric formulation for critical point inflation as a function of $\xi$ and $\kappa$.}
\label{fig:nCritMetricOld}
\end{minipage}
\hfill
\begin{minipage}[t]{.45\textwidth}
\includegraphics[width=\textwidth]{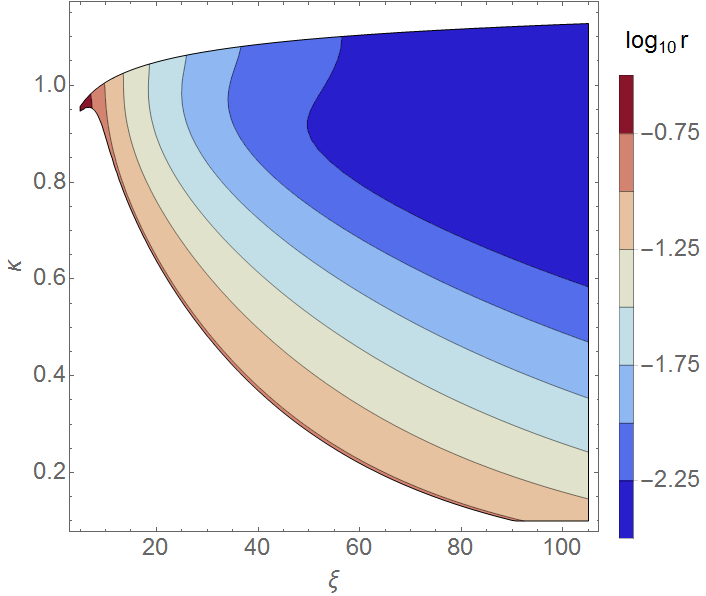}
\caption{Tensor-to-scalar ratio in the metric formulation for critical point inflation as a function of $\xi$ and $\kappa$.}
\label{fig:rCritMetricOld}
\end{minipage}
\\ \vspace{20pt}
\begin{minipage}[t]{.45\textwidth}
\includegraphics[width=\textwidth]{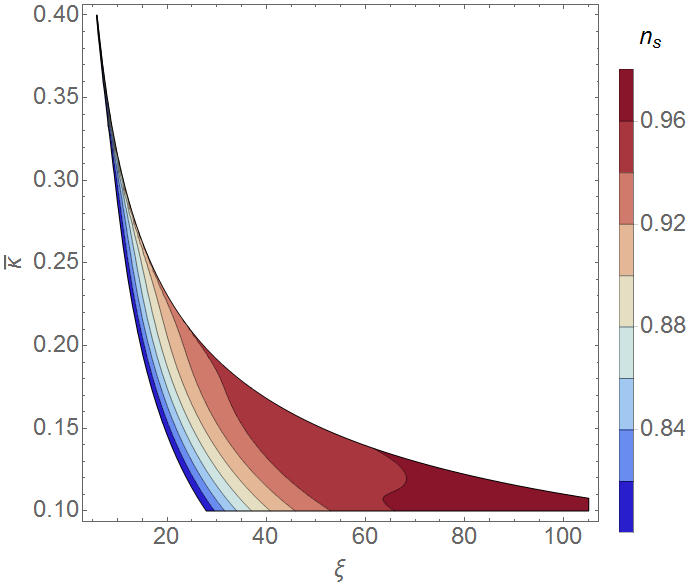}
\caption{Spectral index in the metric formulation for critical point inflation as a function of $\xi$ and $\kappat$.}
\label{fig:nCritMetricNew}
\end{minipage}
\hfill
\begin{minipage}[t]{.45\textwidth}
\includegraphics[width=\textwidth]{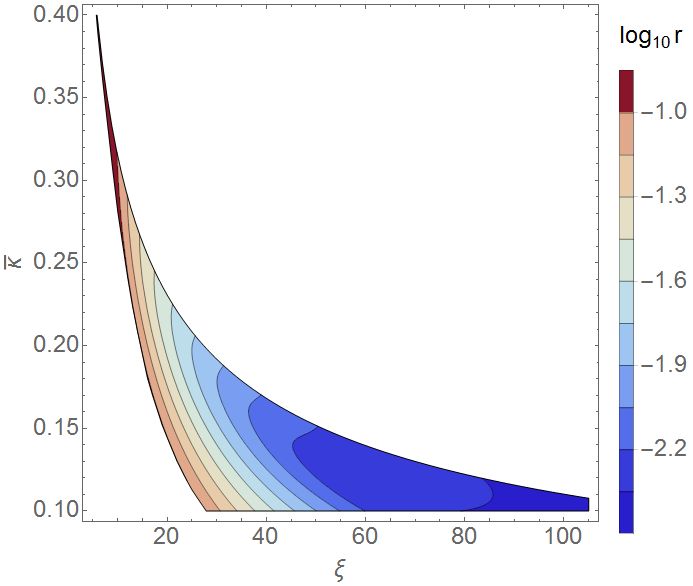}
\caption{Tensor-to-scalar ratio in the metric formulation for critical point inflation as a function of $\xi$ and $\kappat$.}
\label{fig:rCritMetricNew}
\end{minipage}
\end{figure}

\begin{figure}
\begin{minipage}[t]{.45\textwidth}
\includegraphics[width=\textwidth]{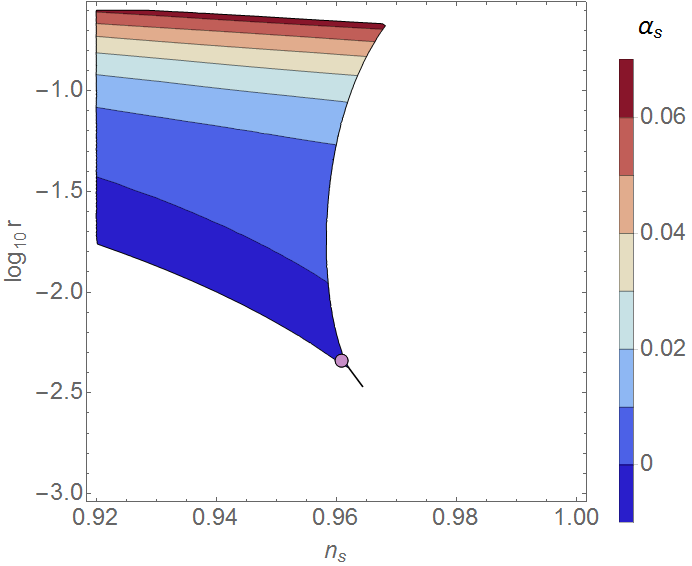}
\caption{Running of the spectral index $\a_s$ in the metric formulation for critical point inflation. The purple dot corresponds to plateau inflation.}
\label{fig:aCritMetric}
\end{minipage}
\hfill
\begin{minipage}[t]{.45\textwidth}
\includegraphics[width=\textwidth]{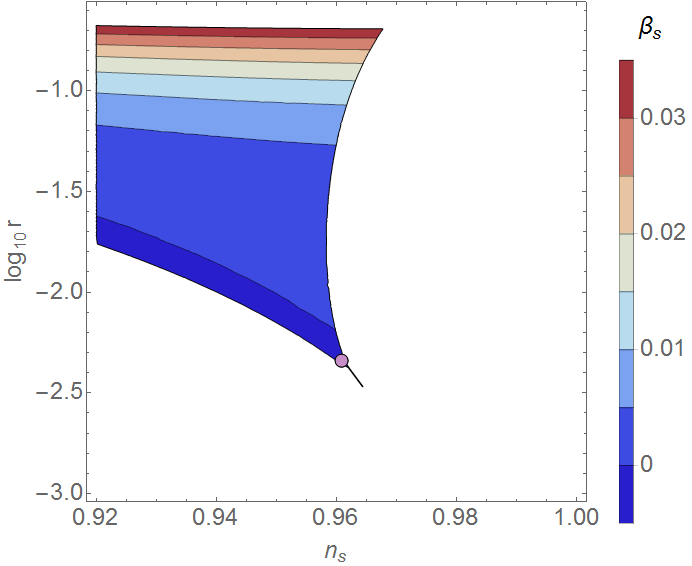}
\caption{Running of the running $\b_s$ in the metric formulation for critical point inflation. The purple dot corresponds to plateau inflation.}
\label{fig:bCritMetric}
\end{minipage}
\end{figure}

Our results in the metric case are presented in \cref{fig:nCritMetricOld,fig:rCritMetricOld,fig:nCritMetricNew,fig:rCritMetricNew,fig:aCritMetric,fig:bCritMetric}. In \cref{fig:nCritMetricOld,fig:rCritMetricOld} we show $n_s$ and $r$ as a function of $\xi$ and $\kappa$. In \cref{fig:nCritMetricNew,fig:rCritMetricNew} we show the case with $\kappat$ instead. We only show values $n_s>0.8$, which gives the bottom boundary. The top boundary comes from the constraint on the number of e-folds.

\Cref{fig:rCritMetricOld,fig:rCritMetricNew} show that the tensor-to-scalar ratio grows with decreasing $\xi$. The largest value of $r$ shown corresponds to $\xi\approx5$. By choosing $\xi$ and $\kappa$ (or $\kappat$) we can adjust $r$ independently from $n_s$ to obtain larger $r$ than in the plateau case, as discussed in \cite{Bezrukov:2014a, Hamada:2014, Bezrukov:2014b, Rubio:2015}.

From \cref{fig:aCritMetric,fig:bCritMetric} we see that large $r$ corresponds to large $\a_s$ and $\b_s$. The boundary on the right corresponds to the constraint on the number of e-folds. If reheating were to last longer than four e-folds, the boundary would move to the left, decreasing $n_s$ but not affecting $r$. The cut-off in the possible values of $\kappa$ (or $\kappat$) has no effect on the plots. The current limits from Planck temperature and low multipole polarisation data are $-0.015<\a_s<0.039$ and $-0.003<\b_s<0.059$ (assuming no tensor modes) \cite{Planck:inflation}. The precise limits depend on the datasets considered \cite{Planck:inflation, Cabass:2016}

The limit on $\a_s$ together with the Planck spectral index limit $0.94<n_s<0.97$ \cite{Planck:inflation} gives $r<0.2$, which is weaker than the direct observational bound on $r$. The observational $\b_s$ constraint does not add extra information. The $n_s$ limit leads to the lower bound $\a_s>-2\times10^{-3}$. For large values $\xi h^2\gg1$, the parameter space reduces to the bottom line of \cref{fig:aCritMetric,fig:bCritMetric}, the one parameter along the line corresponding to the value of $\d_c^2/(\xi^2 h_*^4)$, as discussed in \sec{sec:crit}. The purple dot marks the plateau case, to which the results reduce to for all allowed values of $\kappa$ when $\xi>500$.

Let us compare to the analytical estimates of \sec{sec:crit}. For $\kappa=1$ and $\xi=20$ the numerical calculation gives $r=0.03$ and $n_s=0.92$, which roughly agree with the analytical result. However, this is something of a coincidence, because the field value at the pivot scale $h_*=1.1$ is actually far from the critical point value $h_c=0.28$. This is related to the fact that we did not consider the constraint on the number of e-folds. In the case $\xi h^2\gg1$, we recover the values $n_s=0.96$ and $r=4\times10^{-3}$ with $\xi=70$ and $\kappa=0.8$, practically coinciding with the plateau result. In the case $\kappa=\kappat\sqrt{\xi}$, we instead need $\kappat=0.1$.

Let us then compare to the results of \cite{Hamada:2014, Bezrukov:2014a, Enckell:2016}. We recover the result of \cite{Hamada:2014}, and a detailed comparison shows that we recover the results of \cite{Bezrukov:2014a}, with the exception that our constraint on the number e-folds cuts off values of $n_s\gtrsim0.97$.
Our results for $n_s$, $r$ and $\a_s$ shown in \cref{fig:aCritMetric} agree well with the results in figure 1 of \cite{Enckell:2016}, despite our simpler approximation for the loop corrections. (In \cite{Enckell:2016} the running of the quartic coupling was calculated with Standard Model and chiral Standard Model renormalisation group equations, with a jump as a free parameter to account for the intermediate region.) The main difference is our boundary on the right from the number of e-folds, cutting off values $n_s\gtrsim0.97$.

Let us now see how the results change in the Palatini formulation.

\para{Palatini formulation.}

\begin{figure}[t]
\center
\begin{minipage}[t]{.45\textwidth}
\includegraphics[width=\textwidth]{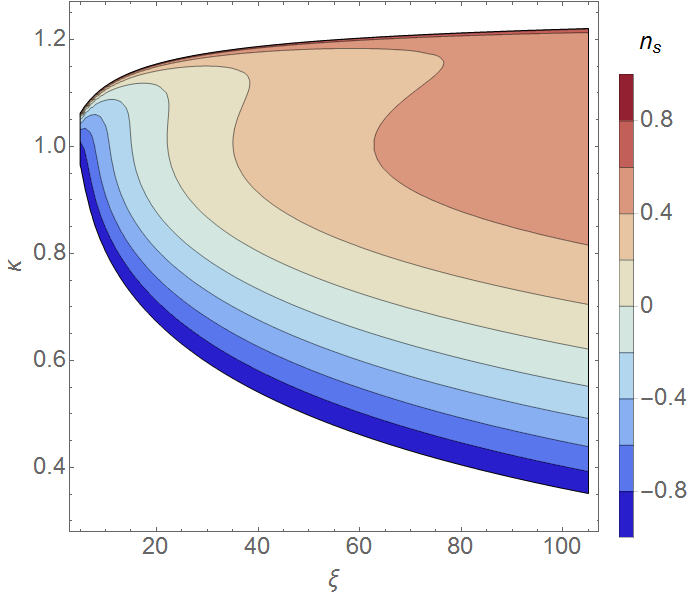}
\caption{Spectral index in the Palatini formulation for critical point inflation as a function of $\xi$ and $\kappa$.}
\label{fig:nCritPalatiniOld}
\end{minipage}
\hfill
\begin{minipage}[t]{.45\textwidth}
\includegraphics[width=\textwidth]{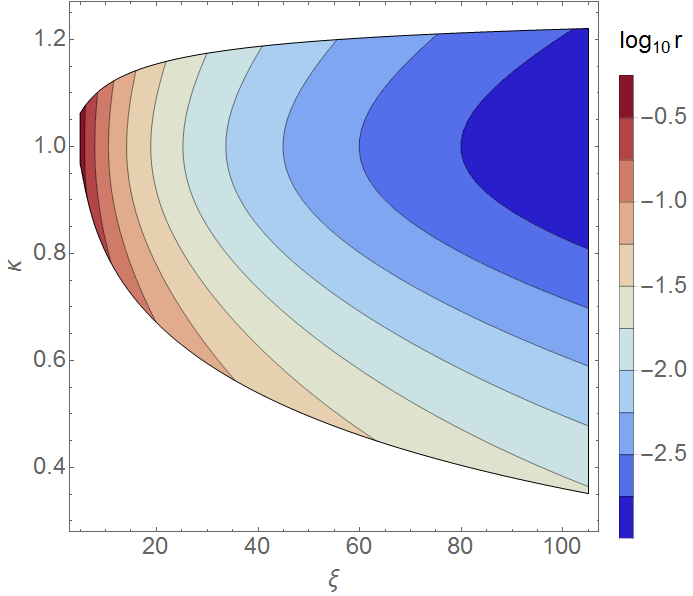}
\caption{Tensor-to-scalar ratio in the Palatini formulation for critical point inflation as a function of $\xi$ and $\kappa$.}
\label{fig:rCritPalatiniOld}
\end{minipage}
\end{figure}

\begin{figure}[t]
\center
\begin{minipage}[t]{.45\textwidth}
\includegraphics[width=\textwidth]{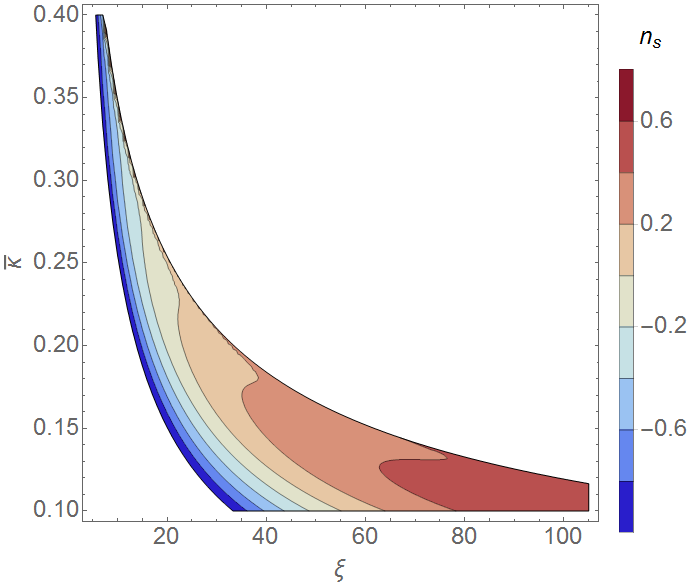}
\caption{Spectral index in the Palatini formulation for critical point inflation as a function of $\xi$ and $\kappat$. (Note that the range for $n_s$ is different than in the metric case.)}
\label{fig:nCritPalatiniNew}
\end{minipage}
\hfill
\begin{minipage}[t]{.45\textwidth}
\includegraphics[width=\textwidth]{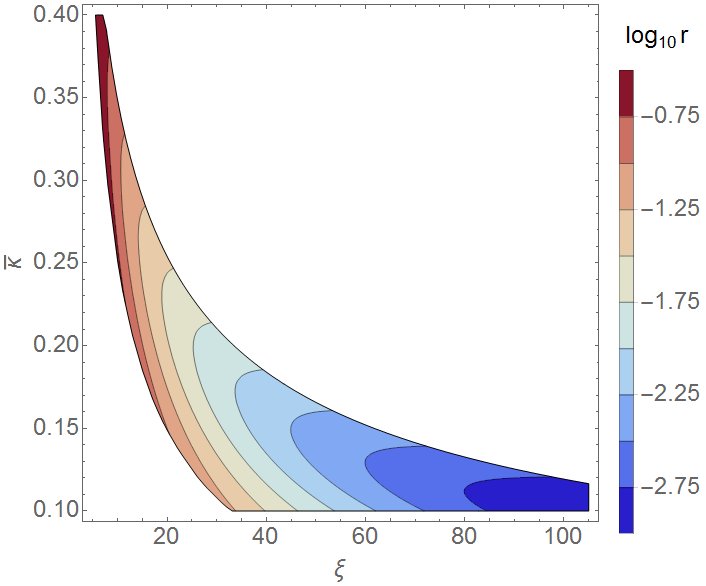}
\caption{Tensor-to-scalar ratio in the Palatini formulation for critical point inflation as a function of $\xi$ and $\kappat$.}
\label{fig:rCritPalatiniNew}
\end{minipage}
\end{figure}

\begin{figure}[t]
\center
\begin{minipage}[t]{.45\textwidth}
\includegraphics[width=\textwidth]{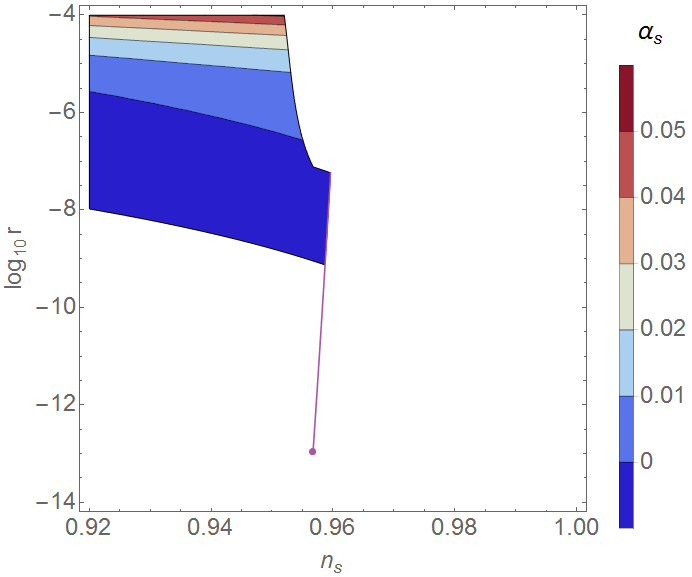}
\caption{Running of the spectral index $\a_s$ in the Palatini formulation for critical point inflation with $\kappa$. The purple line corresponds to $\xi h^2\gg1$, when the results reduce to the plateau case.
}
\label{fig:aCritPalatiniOld}
\end{minipage}
\hfill
\begin{minipage}[t]{.45\textwidth}
\includegraphics[width=\textwidth]{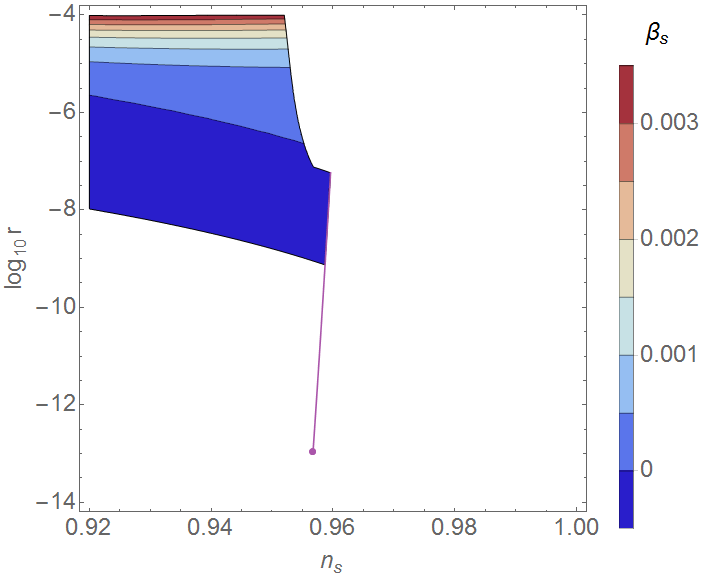}
\caption{Running of the running $\b_s$ in the Palatini formulation for critical point inflation with $\kappa$. The purple line corresponds to $\xi h^2\gg1$, when the results reduce to the plateau case.
}
\label{fig:bCritPalatiniOld}
\end{minipage}
\end{figure}

The Palatini formulation results are shown in \cref{fig:rCritPalatiniOld,fig:nCritPalatiniOld,fig:rCritPalatiniNew,fig:nCritPalatiniNew,fig:aCritPalatiniOld,fig:bCritPalatiniOld}.
In \cref{fig:rCritPalatiniOld,fig:nCritPalatiniOld} we plot $n_s$ and $r$ as a function of $\xi$ and $\kappa$.
In \cref{fig:rCritPalatiniNew,fig:nCritPalatiniNew} we use $\kappat$ instead of $\kappa$. Unlike in the metric case, the observationally allowed range is just a small sliver at the top of the plot, bounded from above by the constraint on the number of e-folds. (We have cut off the plot at $n_s<-1$ for clarity.) 
\Cref{fig:rCritPalatiniOld,fig:rCritPalatiniNew} show that the tensor-to-scalar ratio decreases with growing $\xi$, as in the metric case. The spectral index decreases as $\kappa$ and $\kappat$ decrease, as \cref{fig:nCritPalatiniOld,fig:nCritPalatiniNew} show. As in the metric case, we can tune $\kappa$ (or $\kappat$) and $\xi$ in this region to match $r$ and $n_s$ with observations. We could obtain tensor-to-scalar ratio as large as $0.1$, were it not for the constraints on the running of $n_s$.

When we use $\kappa$, as in \cite{Bezrukov:2014a, Bezrukov:2017}, the results reduce to the plateau case for $\xi\gtrsim\times10^6$. When we use $\kappat$, following \cite{Hamada:2014}, there are no solutions with $\kappa>0.1$ for $200<\xi<6\times10^6$, as inflation does not last long enough. For $\xi>6\times10^6$ the results reduce to the plateau case.

In \cref{fig:aCritPalatiniOld,fig:bCritPalatiniOld} we show $\a_s$ and $\b_s$ in the case with $\kappa$ as a function of $n_s$ and $r$.
As in the metric case, the boundary on the upper right comes from the constraint on the number of e-folds, and changing the duration of reheating shifts it horizontally, affecting the allowed values of $n_s$, without much effect on $r$. Unlike in the metric case, in Palatini case, the values of $\kappa$ have a drastic effect on the allowed parameter region. The bottom boundary comes from our (somewhat arbitrary) cut-off $\kappa>0.1$; decreasing the cut-off would move the boundary down.
\Cref{fig:aCritPalatiniOld,fig:bCritPalatiniOld} show that large $r$ leads to large running, as in the metric case, but because $r$ is suppressed, the Planck constraint $\a_s<0.039$ is more important. Taking also the Planck constraint on the spectral index into account reduces the allowed range of $r$ to  $1\times10^{-13}<r<7\times10^{-5}$. The maximum value corresponds to $\xi=540$. This range is sensitive to the precise observational limit on $n_s$ and the closely related number of e-folds: increasing the duration of reheating from $\Delta N_\reh=4$ would further reduce $r$.

The running of the running is suppressed, we have $-5\times10^{-5}<\b_s<0.059$ after the $n_s$ constraint is imposed, compared to the metric case range $-2\times10^{-4}<\b_s<0.059$. The Palatini formulation does not allow $r$ as large as the metric formulation on the plateau (let alone at the critical point). The one-dimensional purple line extending to small $r$ in \cref{fig:aCritPalatiniOld,fig:bCritPalatiniOld} corresponds to the fact that for $\xi h^2\gg1$ (in this case corresponding to $\xi>10^6$) the results are independent of $\kappa$, there is only one free parameter and we recover the plateau results.

In the case of $\kappat$, the limit $\kappat>0.1$ essentially removes the possibility of critical point inflation with spectral index and running compatible with observations, except when the situation reduces to the plateau case.

As in the metric case, the analytical estimates of \sec{sec:crit} are correct only for $\xi h^2\gg1$. The reason is again that the field value at the pivot scale is far from the critical point value, because we did not consider the constraint on the number of e-folds. For the case $\kappa=\kappat\sqrt{\xi}$, we would need the too small value $\kappat=8\times10^{-3}$, in agreement with the estimate in \sec{sec:crit}.

\subsection{Hilltop inflation}

\para{Metric formulation.}

\begin{figure}[t]
\center
\begin{minipage}[t]{.45\textwidth}
\includegraphics[width=\textwidth]{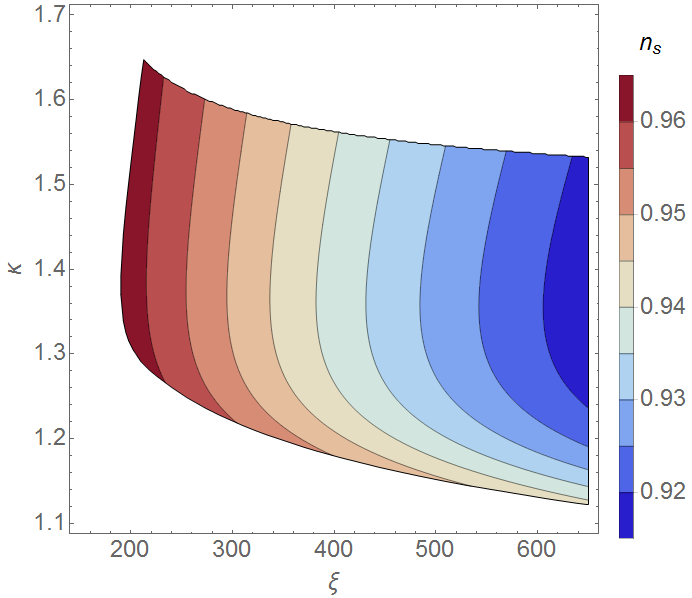}
\caption{Spectral index in the metric formulation in hilltop inflation as a function of $\xi$ and $\kappa$.}
\label{fig:nHillMetricOld}
\end{minipage}
\hfill
\begin{minipage}[t]{.45\textwidth}
\includegraphics[width=\textwidth]{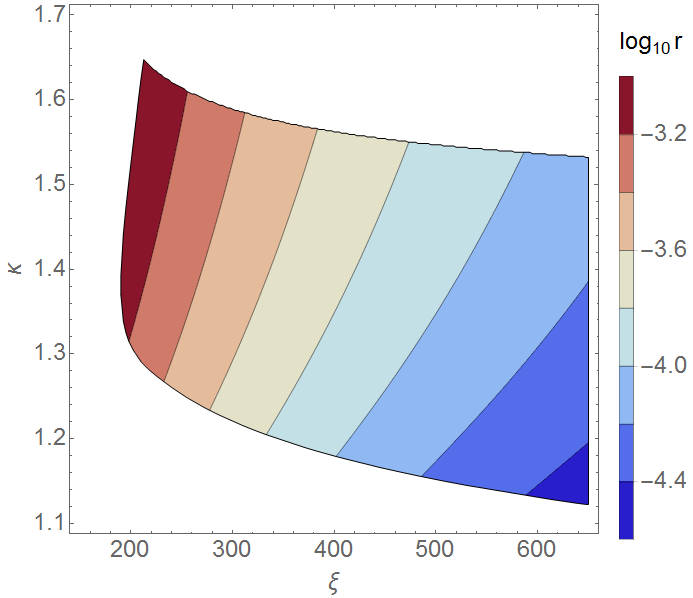}
\caption{Tensor-to-scalar ratio in the metric formulation in hilltop inflation as a function of $\xi$ and $\kappa$.}
\label{fig:rHillMetricOld}
\end{minipage}
\\ \vspace{20pt}
\begin{minipage}[t]{.45\textwidth}
\includegraphics[width=\textwidth]{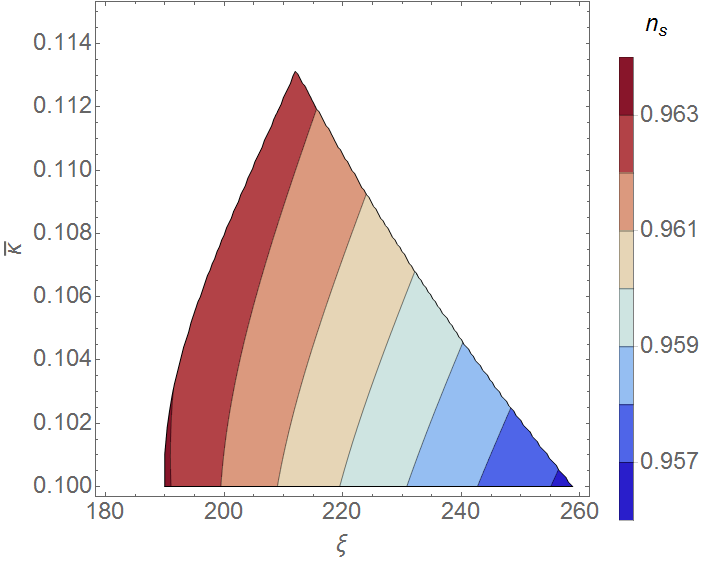}
\caption{Spectral index in the metric formulation in hilltop inflation as a function of $\xi$ and $\kappat$.}
\label{fig:nHillMetricNew}
\end{minipage}
\hfill
\begin{minipage}[t]{.45\textwidth}
\includegraphics[width=\textwidth]{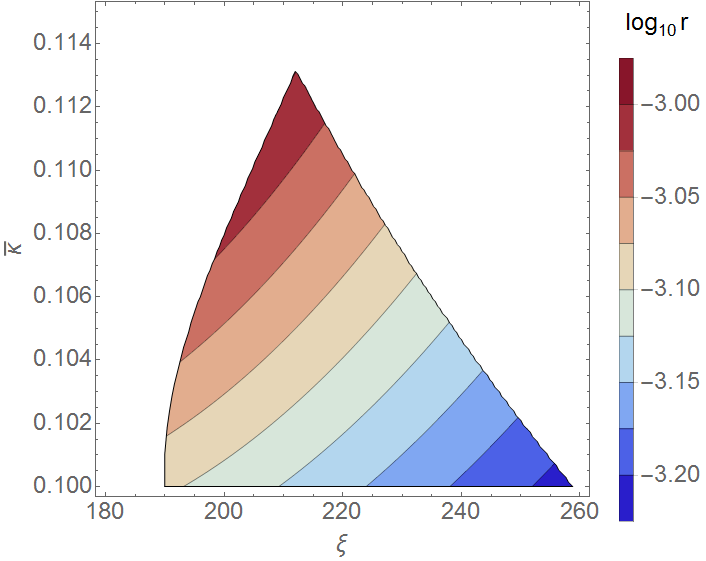}
\caption{Tensor-to-scalar ratio in the metric formulation in hilltop inflation as a function of $\xi$ and $\kappat$.}
\label{fig:rHillMetricNew}
\end{minipage}
\end{figure}

\begin{figure}
\center
\begin{minipage}[t]{.45\textwidth}
\includegraphics[width=\textwidth]{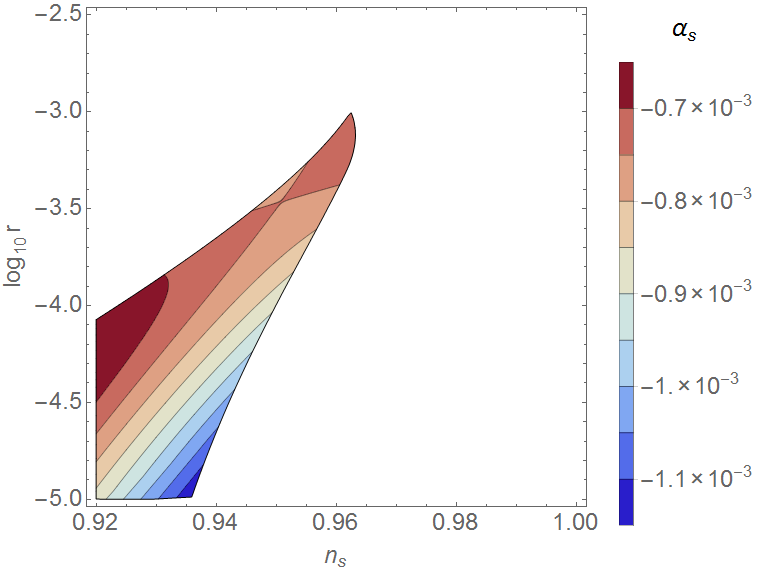}
\caption{Running of the spectral index $\a_s$ in the metric formulation for hilltop point inflation with $\kappa$.}
\label{fig:abHillMetricOld}
\end{minipage}
\hfill
\begin{minipage}[t]{.45\textwidth}
\includegraphics[width=\textwidth]{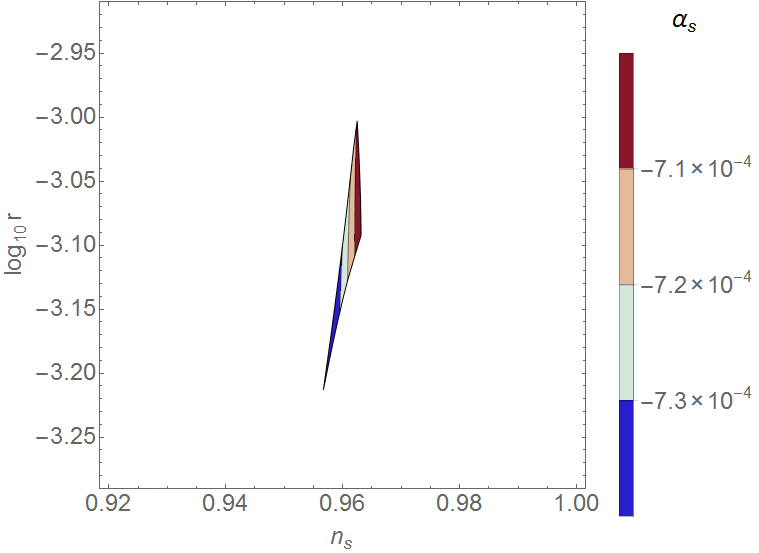}
\caption{Running of the spectral index $\a_s$ in the metric formulation for hilltop point inflation with $\kappat$.}
\label{fig:abHillMetricNew}
\end{minipage}
\end{figure}

Our results for inflation at the hilltop in the metric formulation are shown in \cref{fig:rHillMetricOld,fig:nHillMetricOld,fig:rHillMetricNew,fig:nHillMetricNew,fig:abHillMetricOld,fig:abHillMetricNew}. In \cref{fig:rHillMetricOld,fig:nHillMetricOld} we plot $n_s$ and $r$ as a function of $\xi$ and $\kappa$. For $\xi<215$, the boundary on the left and bottom is determined by the upper bound $\lambda_0<b/16$ and the existence of the hilltop. For larger $\xi$, the bottom boundary is determined by the constraint on e-folds. The top boundary comes from $\lambda_0>0$. We see that $r$ and $n_s$ grow when $\xi$ decreases or $\kappa$ (or $\kappat$) grows; the behaviour of $n_s$ is opposite to the critical point case, where it goes down with decreasing $\xi$. However, unlike in the critical point case, $r$ is suppressed, not enhanced, with regard to inflation on the plateau.
The trends are the same for the $\kappat$ case, shown in \cref{fig:rHillMetricNew,fig:nHillMetricNew}. The left boundary comes from upper bound $\lambda_0<b/16$ and the existence of the hilltop, and the right boundary comes from the cut-off $\lambda_0>0$.

In \cref{fig:abHillMetricOld} we show the allowed region in the $(n_s,r)$ plane for the $\kappa$ case. The right boundary is the same as the left and bottom boundary in \cref{fig:rHillMetricOld,fig:nHillMetricOld}, and the left boundary comes from $\l_0>0$. Unlike in critical point inflation, the maximum value of $r$ is below the plateau case; taking into account the Planck constraint on $n_s$ and $\a_s$, we have $2\times10^{-5}<r<1\times10^{-3}$. Both $\a_s$ and $\b_s$ are well below the Planck upper limits and almost constant; accounting for the Planck constraint on $n_s$, we have $-1\times10^{-3}<\a_s<-7\times10^{-4}$ and $-3\times10^{-4}<\b_s<-1\times10^{-4}$.

\Cref{fig:abHillMetricNew} shows the allowed region in the $\kappat$ case, which is now reduced to a small sliver, with values in agreement with observations. The top-right boundary is the same as the left-side boundary in \cref{fig:rHillMetricNew,fig:nHillMetricNew}. The bottom right boundary comes from the cut-off $\kappat>0.1$ and the left boundary from the cut-off $\lambda_0>0$. Again, the running and running of the running are small, $\a_s=-7\times10^{-4}$, $\b_s=-(1\ldots2)\times10^{-4}$.

Changing the duration of reheating has a more complicated effect than in critical point inflation. In the $\kappa$ case the main effect of longer reheating would be to slightly decrease the allowed value for $n_s$, keeping it within the observationally allowed region. In the $\kappat$ case, the maximum value of $r$ would grow by about a factor of 2 for $\Delta N_\reh=10$.

Let us briefly compare these results with the ones given in table 1 of \cite{Fumagalli:2016}, where the loop corrections were evolved with renormalisation group running, not applying threshold corrections. The values are compatible with the $\kappa$ case shown in \cref{fig:abHillMetricOld}, and show the same trend that large $n_s$ values correspond to large values of $r$. They are not compatible with the $\kappat$ case due to our cut-offs $\lambda_0>0$ and $\kappa>0.1$. We remind that our approximation for the loop corrections may not be accurate in the hilltop case.

\para{Palatini formulation.}

\begin{figure}[t]
\center
\begin{minipage}[t]{.45\textwidth}
\includegraphics[width=\textwidth]{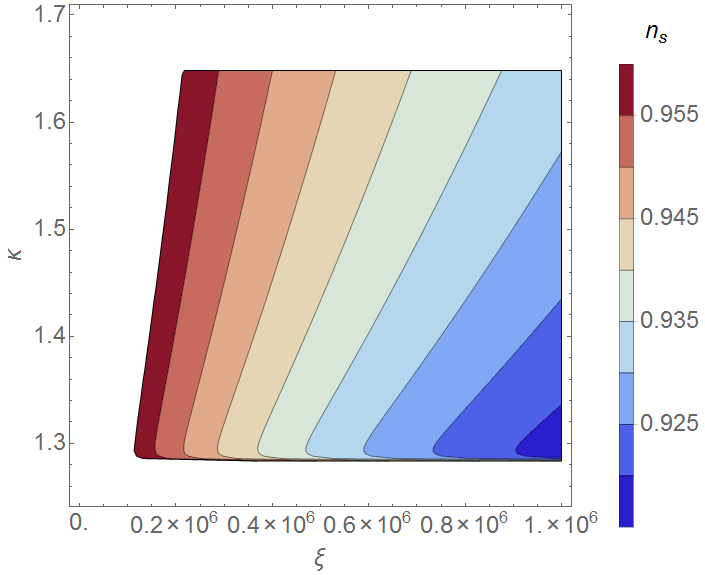}
\caption{Spectral index in the Palatini formulation in hilltop inflation as a function of $\xi$ and $\kappa$.}
\label{fig:nHillPalatiniOld}
\end{minipage}
\hfill
\begin{minipage}[t]{.45\textwidth}
\includegraphics[width=\textwidth]{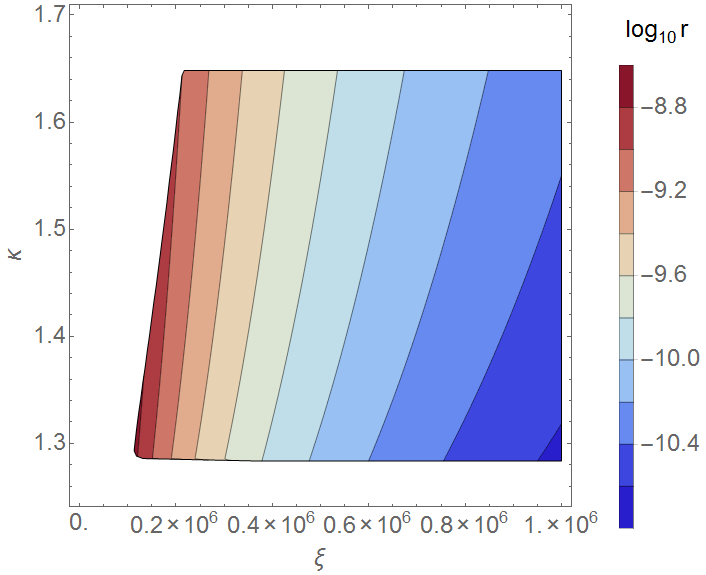}
\caption{Tensor-to-scalar ratio in the Palatini formulation in hilltop inflation as a function of $\xi$ and $\kappa$.}
\label{fig:rHillPalatiniOld}
\end{minipage}
\end{figure}

\begin{wrapfigure}{R}{0.5\textwidth}
\centering
\begin{minipage}[t]{.45\textwidth}
\vspace{-20pt}
\includegraphics[width=\textwidth]{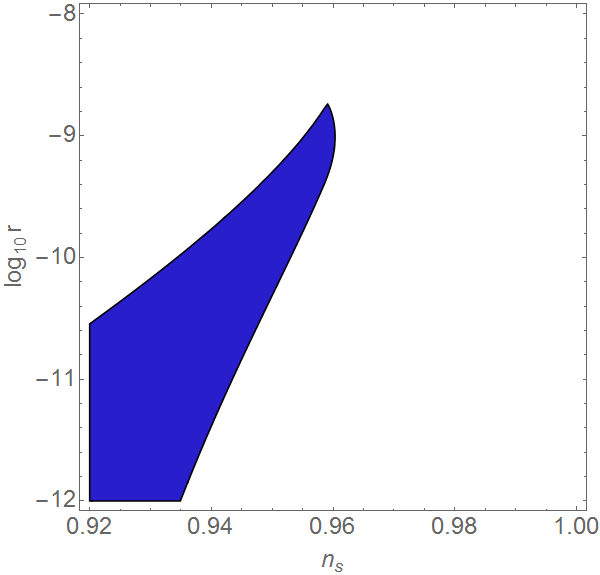}
\caption{Allowed parameter region on the $(n_s,r)$ plane in the Palatini formulation for Hilltop inflation with $\kappa$. The running and the running of the running are not single-valued on this plane.}
\label{fig:abHillPalatiniOld}
\end{minipage}
\vspace{-30pt}
\end{wrapfigure}

Our results in the Palatini formulation are shown in \cref{fig:rHillPalatiniOld,fig:nHillPalatiniOld,fig:abHillPalatiniOld}. In \cref{fig:rHillPalatiniOld,fig:nHillPalatiniOld} we plot $n_s$ and $r$ as a function of $\xi$ and $\kappa$. The top boundary results from the cut-off $\lambda_0 > 0$. For $\xi>2\times10^5$ the bottom boundary is determined by the constraint on e-folds. Otherwise the boundary results from the upper bound of $\lambda_0<b/16$ and the existence of the hilltop. As in the metric case, $n_s$ and $r$ grow with increasing $\xi$ and, to a lesser extent, $\kappa$. Unlike in critical point inflation, where the required values of $\xi$ are similar, at the hilltop the Palatini case requires a much larger non-minimal coupling than the metric case, $\xi>2\times10^5$, and $r$ is even more suppressed. It is possible to get the same $n_s$ and $r$ for multiple pairs of $\xi$ and $\kappa$.

In \cref{fig:abHillPalatiniOld} we show the allowed region on the $(n_s,r)$ plane. After accounting for the Planck limit on $n_s$, we have $1\times10^{-10}<r<2\times10^{-9}$.
We do not show $\a_s$ or $\b_s$, as their projection onto this plane is not single-valued, and the variations are small. After accounting for the Planck constraint on $n_s$, we get $\a_s=-(8\ldots9)\times10^{-4}$ and $\b_s=-(2\ldots3)\times10^{-5}$, the latter being suppressed by one order of magnitude compared to the metric case. Both are well below current observational limits.
The right boundary corresponds to the left and bottom boundaries in \cref{fig:rHillPalatiniOld,fig:nHillPalatiniOld}. The left boundary comes from the fact that for a constant $r$ there is a minimum for $n_s$ as can be seen from \cref{fig:nHillPalatiniOld,fig:rHillPalatiniOld}. Increasing $\Delta N_\reh$ has a similar effect as in the metric case, lowering $n_s$ and slightly increasing $r$, which however remains suppressed.

In the case of $\kappat$, the limit $\kappat>0.1$ removes the possibility of hilltop inflation with spectral index and running compatible with observations.

\section{Conclusions} \label{sec:conc}

\para{Inflationary signatures of gravitational degrees of freedom.}

We have studied the differences in the Higgs inflation predictions in the metric and Palatini formulations of general relativity. We have used a simple approximation for loop corrections to get analytical understanding in addition to numerical results, considering the two different parametrisations of \cite{Bezrukov:2014a, Bezrukov:2017} and \cite{Hamada:2014}. We have considered inflation on the plateau and away from there, in particular at a critical point, at a hilltop and in a false vacuum. The tensor-to-scalar ratio is different for inflation in different parts of the potential, and Higgs inflation outside the plateau has been considered particularly for the possibility of obtaining a tensor-to-scalar ratio larger than the plateau result $r=5\times10^{-3}$.

In the metric case, inflation at the critical point can enhance the tensor amplitude from the plateau value up to $r=0.2$, while in hilltop inflation it is suppressed to $2\times10^{-5}<r<1\times10^{-3}$, after accounting for Planck observational constraints on the spectral index $n_s$ and the running $\a_s$. The running cannot be strongly negative in either case, $\a_s>-2\times10^{-3}$. These metric case results agree with previous work, when the different approximation schemes are taken into account \cite{Hamada:2014, Bezrukov:2014a, Enckell:2016, Fumagalli:2016}.

In the Palatini formulation, we find that both critical point and hilltop inflation are viable in the parametrisation of \cite{Bezrukov:2014a, Bezrukov:2017}, but not in that of \cite{Hamada:2014}. (For false vacuum inflation, there is no major difference between the metric and Palatini formulations.) As in the plateau case studied in \cite{Bauer:2008}, $r$ is highly suppressed for the Palatini formulation, with the value \mbox{$1\times10^{-13}<r<7\times10^{-5}$} in the critical point case and $1\times10^{-10}<r<2\times 10^{-9}$ at the hilltop, once observational constraints on the spectral index and its running are taken into account. (For large values of the non-minimal coupling $\xi$, the results approach the tree level plateau case, in agreement with general arguments \cite{Fumagalli:2016sof}.) These values are sensitive to the precise observational value of the spectral index and to changes in reheating. For the $r$ (and $n_s$) values that overlap between the metric and Palatini cases, the running and the running of the running set them apart. So if Higgs is the inflaton, observations of the cosmic microwave background and large-scale structure can be used to determine whether the metric or the Palatini formulation is correct.

The precise numbers depend on our treatment of the loop corrections. In particular, the approximation we have used may not be valid at the hilltop. It should also be checked whether the parametrisation of \cite{Bezrukov:2014a, Bezrukov:2017} or \cite{Hamada:2014} (or neither) is more appropriate.
Following the renormalisation group equation running in detail would give more accurate results, provided the effects of the mid-field regime $\xi h^2\sim1$ where the loop corrections are not under control is properly parametrised \cite{Bezrukov:2014a, Bezrukov:2014b, Rubio:2015, George:2015, Saltas:2015, Bezrukov:2010, Salvio:2013, George:2013, Enckell:2016, Fumagalli:2016, Bezrukov:2017, Fumagalli:2016sof}. 
Inflation in the mid-field regime $\xi h^2\sim1$ is particularly sensitive to the unknown corrections to the shape of the potential, but even the predictions of the large field regime $\xi h^2\gg1$ can be modified through a change in the total number of e-folds due to possible features, such as a critical point.

The possible dependence on the particle physics UV completion has been much discussed in the literature \cite{Bezrukov:2014a, Bezrukov:2014b, Rubio:2015, George:2015, Saltas:2015, George:2013, Calmet:2013, Barvinsky:2009, Burgess:2009,  Bezrukov:2010, Bezrukov:2012, Enckell:2016, Fumagalli:2016, Bezrukov:2017, Fumagalli:2016sof}. We have focused on a less appreciated ambiguity: the predictions of Higgs inflation are also sensitive to the choice of the gravitational degrees of freedom (presumably ultimately determined by the UV completion of the gravity sector). Uncertainties of the quantum corrections notwithstanding, the suppression of the tensor-to-scalar ratio is a robust signature of the Palatini formulation, originating in the different field transformation to obtain a canonical kinetic term for the Einstein frame Higgs field in the metric and Palatini formulations, tracing to the different structure of the Ricci tensor in the two cases.

If $r$ is detected at a level that is within the reach of next generation experiments, the Palatini case will be ruled out, if our approximative treatment is valid. Turning this around, if $r$ at the level predicted by the metric formulation $r>2\times10^{-5}$ is not seen, this will not rule out Higgs inflation, only Higgs inflation in the metric formulation (with caveats about the treatment of the quantum corrections). In the context of Higgs inflation, inflationary predictions can thus distinguish between different gravitational degrees of freedom. Such tests can also be extended to other formulations of general relativity that are equivalent for the Einstein--Hilbert action plus a minimally coupled scalar field, but distinct in Higgs inflation, where non-minimal coupling plays a central role.

\acknowledgments

We thank Fedor Bezrukov and Dani Figueroa for helpful discussions and correspondence. PW is supported by Finnish Cultural Foundation, Kymenlaakso Regional fund.

\printbibliography

\end{document}